\documentclass[preprint,5p,times]{elsarticle} 
\usepackage[final]{microtype}     
\usepackage{siunitx}              
\usepackage{indentfirst}          
\usepackage{hyperref}
\usepackage{enumitem}
\usepackage{amsmath}
\usepackage{booktabs} 
\setlist{leftmargin=*, itemsep=2pt, topsep=2pt, parsep=0pt, partopsep=0pt}
\usepackage{booktabs}

\journal{Physics of the Dark Universe}

\begin{document}

\begin{frontmatter}

\title{Probing low-mass dark matter from sub-MeV to sub-GeV with germanium-based quantum phononic spectroscopy}

\author[usd]{D.-M.\ Mei\corref{cor1}}
\author[usd]{N.\ Budhathoki}
\author[usd]{S.\ A.\ Panamaldeniya}
\author[usd]{K.-M.\ Dong}
\author[usd]{S.\ Bhattarai}
\author[usd]{A.\ Warren}
\author[usd]{A.\ Prem}
\author[usd]{S.\ Chhetri}
\address[usd]{Department of Physics, University of South Dakota, Vermillion, SD 57069, USA}
\cortext[cor1]{Corresponding author. Email: \texttt{dongming.mei@usd.edu}}

\begin{abstract}
We present a germanium phonon-to-charge transducer that integrates a slow-phonon phononic-crystal (PnC) region with radio-frequency quantum point-contact (RF-QPC) readout at \SI{4}{K}, and we evaluate its dark-sector reach. A calibrated signal-collection model—combining geometric guiding, propagation survival, and multiplicity-assisted primary-phonon detection—provides selection-corrected thresholds in the $10^{-3}$–$10^{-2}$~eV range and a background model informed by nanosecond timing gates and GHz-band power–spectral–density windows. Under standard halo assumptions, a \SI{100}{g} module achieves projected sensitivity to DM–electron and DM–nucleon scattering at recoil energies below $10^{-2}$~eV, probing cross sections below $10^{-43}\,\mathrm{cm}^{2}$ for $m_\chi \in [0.01,100]~\mathrm{MeV}/c^{2}$ (with efficiencies and thresholds folded in). We present sensitivities for both heavy-mediator ($F_{\rm DM}=1$) and light-mediator ($F_{\rm DM}\propto 1/q^{2}$) benchmarks, quantify dominant systematics (threshold, phonon quality factor $Q$, and bulk defect densities), and outline a staged program toward kg$\cdot$yr exposures that begins to test models approaching the solar CE$\nu$NS background.
\end{abstract}


\end{frontmatter}

\section{Introduction} Low-mass dark matter (DM) and solar neutrinos motivate detector concepts with effective thresholds \emph{sub-eV} on the gram-to-kilogram scale. In this work, we present a germanium (Ge)-based quantum sensor that converts guided, athermal phonons into charge signals and evaluate its dark-sector reach from \emph{sub-MeV} to \emph{sub-GeV} masses. Solar coherent elastic neutrino–nucleus scattering (CE$\nu$NS) emerges naturally as both a physics by-product and an irreducible background in the same energy range~\cite{Freedman1974,Billard2014}. 

Existing low-threshold programs have made rapid progress, yet their sensitivity is ultimately limited by charge/scintillation thresholds of $\mathcal{O}(10^{-1}$--$10^{3})$~eV, leaving sizable regions of sub-MeV DM parameter space weakly constrained~\cite{Essig2012,Battaglieri2017,Aprile2023,LZ2023,Agnese2018}. We address this gap with a Ge-based quantum sensor for low-energy physics (\emph{GeQuLEP}~\cite{Mei2025}) in which dipole-defined quantum wells (QWs) in ultrapure Ge couple to a slow-phonon region of the phononic crystal (PnC) and a radio frequency quantum point-contact (RF-QPC) for charge readout at cryogenic temperature. The PnC steers and concentrates ballistic phonons from rare-event energy deposits into the sensing region, whereas the RF-QPC transduces lattice-driven charge displacement without requiring bulk charge drift or metal contacts across the active volume. 

The design targets single-primary-phonon-level sensitivity and scalable contact-minimized phonon spectroscopy suitable for rare event searches. Our analysis frames this device as a \emph{phonon-to-charge transducer} for dark-sector physics. Using a calibrated efficiency model, geometric collection, propagation survival in bulk Ge, and multiplicity-assisted primary-phonon detection, we derive selection-corrected thresholds and backgrounds with nanosecond timing gates and GHz-band power-spectral-density (PSD) windows. Under standard halo assumptions, a 100~g module at 4~K reaches projected selection thresholds in the $10^{-3}$--$10^{-2}$~ eV band and achieves sensitivity to DM--electron and DM--nucleon scattering over $m_\chi\!\in\![\mathrm{sub\text{-}MeV},\mathrm{sub\text{-}GeV}]$, with solar CE$\nu$NS setting the ultimate background floor at higher exposures~\cite{Billard2014}. 

\paragraph{Contributions} (i) We introduce a Ge phonon-to-charge sensor that leverages PnC slow-phonon engineering and RF-QPC readout, eliminating bulk charge transport while preserving strong deformation-potential (DP) coupling. (ii) We construct a transparent efficiency pipeline from phonon generation to induced charge, enabling selection-corrected thresholds and a background model compatible with timing/PSD discrimination. (iii) We present projected DM--electron and DM--nucleon sensitivity curves, open a large previously unexplored region of parameter space, and quantify the dominant systematics—threshold, phonon quality factor $Q$, and defect densities—while outlining a staged roadmap toward kg$\cdot$yr exposures where the solar CE$\nu$NS background becomes relevant.

\paragraph{Organization} Section~\ref{sec:concept_design} introduces the sensor concept and the layout of the device. Section~\ref{sec:phonon_physics} describes primary-phonon generation, anharmonic down-conversion, and ballistic transport in Ge. Section~\ref{sec:reflectivity} analyzes acoustic reflectivity and guidance at the Ge--vacuum boundary and their implications for phonon confinement. Section~\ref{sec:collection} formalizes the phonon-collection model and the primary-phonon detection probability. Section~\ref{sec:readout} develops phonon-to-charge transduction and the RF-QPC induced-charge response. Section~\ref{sec:prototype_backgrounds} details the prototype geometry and dominant backgrounds. Section~\ref{sec:timing_psd} defines the timing gate and GHz-band PSD selection windows used for background rejection. Section~\ref{sec:projected_sensitivity} presents projected sensitivity to low-mass dark matter and discusses the solar CE$\nu$NS background. Systematics and calibration are treated in Section~\ref{sec:systematics_calibration}, and the experimental roadmap with risk mitigation appears in Section~\ref{sec:roadmap}. The discussion and outlook are given in Section~\ref{sec:discussion}. We conclude in Section~\ref{sec:conclusion}. 

\section{Concept and Design}
\label{sec:concept_design}

We propose a Ge phonon--to--charge architecture (\emph{GeQuLEP}~\cite{Mei2025}) that combines
(i) localized charge states (``quantum dots'') in ultra–high–purity Ge,
(ii) a near–surface PnC that slows and guides athermal phonons, and
(iii) an RF-QPC for charge readout at cryogenic temperature.
The goal is single–primary–phonon sensitivity with scalable, contact–minimized phonon spectroscopy for rare–event searches.

\subsection{Localized charge states in Ge: gate-defined and impurity-bound}

At cryogenic temperatures, Ge supports two practical routes to quantum dots (QDs):
\emph{(1) gate-defined} electrostatic confinement using lithographic electrodes, and
\emph{(2) impurity-bound} localization when shallow dopants \emph{freeze out} below
$\sim\!\SI{10}{K}$.\footnote{We use ``quantum dot'' for any localized bound state with discrete levels and strong deformation–potential coupling.}
Both mechanisms yield compact wavefunctions that couple efficiently to acoustic phonons.

\paragraph{Gate-defined dots}
Voltages on patterned gates reshape local band edges to form a smooth confining potential well for electrons or holes~\cite{Kouwenhoven1997,Hanson2007}.
For layout and analytics we adopt a Gaussian ansatz~\cite{Burkard1999}:
\begin{equation}
  V_{\mathrm{gate}}(x) \;=\; -V_0 \exp\!\left[-\frac{x^{2}}{2\sigma^{2}}\right],
  \label{eq:Vgate}
\end{equation}
with depth $V_0$ and width $\sigma$ setting dot size and level spacing.

\paragraph{Impurity-bound dots and dipole states below \texorpdfstring{$\mathbf{10}$}{10}\,K}
In ultra–pure Ge operated at $T \lesssim \SI{10}{K}$, shallow donors (e.g. phosphorus (P)) and acceptors (e.g. boron (B), aluminum (Al), gallium (Ga)) form neutral, hydrogenic bound states that localize carriers~\cite{Mei2024,Mei2022}.
A convenient order–of–magnitude model treats the localized pair as an effective electric dipole of size $x$ stabilized when its Coulomb energy balances thermal energy:
\begin{equation}
  x \;\approx\; \frac{q^{2}}{4\pi \varepsilon_{0}\varepsilon_{r} k_{B} T},
  \label{eq:dipole_size}
\end{equation}
with $\varepsilon_{r}\!\approx\!16$ for Ge.
Along the perpendicular bisector of the dipole, an idealized potential reads
\begin{equation}
  V_{\mathrm{dipole}}(x) \;=\; -\,\frac{q^{2}}{4\pi \varepsilon_{0}\varepsilon_{r}\, x},
  \label{eq:Vdipole}
\end{equation}
which we approximate for device modeling by a narrow Gaussian well
\begin{equation}
  V_{\mathrm{imp}}(x) \;\approx\; -V_0' \exp\!\left[-\frac{x^{2}}{2{\sigma'}^{2}}\right],
  \label{eq:Vimp}
\end{equation}
with $V_0' \sim \mathcal{O}(10~\mathrm{meV})$ and $\sigma'$ in the few–nm range at cryogenic temperature (to be adapted to the data)~\cite{san,math}.
This Gaussian form is a standard first–order approximation for analytic estimates of spin / charge phonon interactions~\cite{Kloeffel2012,Maier2015}.

\paragraph{Deformation potential coupling and strain tensor}
The local lattice strain $\varepsilon$ shifts the band edges and the dot levels through the deformation potential (DP) interaction~\cite{BirPikus1974,Mahan2000}.
To leading (dilational) order,
\begin{equation}
  \Delta E_{\mathrm{DP}} \;=\; D \,\mathrm{Tr}\,(\varepsilon),
  \label{eq:DP_shift}
\end{equation}
where $D \approx \SI{13.5}{eV}$\ \text{for electrons in Ge}~\cite{Jacoboni1983},
$\mathrm{Tr}\,(\varepsilon$) is the volumetric (hydrostatic) strain.
In 1D, $\varepsilon(x)=\partial u/\partial x$ for displacement field $u(x)$.
In 3D, $\varepsilon_{ij}=\frac{1}{2}(\partial u_i/\partial x_j + \partial u_j/\partial x_i)$
and $\mathrm{Tr}\,(\varepsilon)=\varepsilon_{xx}+\varepsilon_{yy}+\varepsilon_{zz}$~\cite{Cleland2003}.

\begin{figure}[htp]
  \centering
  \includegraphics[width=0.85\linewidth]{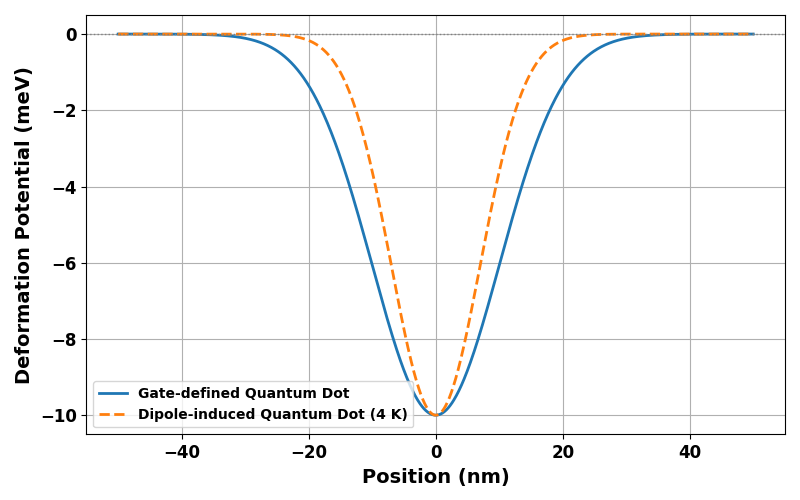}
  \caption{%
  Comparison of DPs in Ge:
  a gate-defined Gaussian well (solid blue) and an impurity-induced, dipole-bound Gaussian well at \SI{4}{K} (dashed orange). The Gaussian widths are $\sigma=\mathbf{16.2}\,\mathrm{nm}$ (gate-defined) and $\sigma'=\mathbf{12.4}\,\mathrm{nm}$ (impurity-induced). These values were chosen to reproduce the level spacing and localization length expected for our lithographic gate geometry at \(\mathbf{4}\,\mathrm{K}\) and acceptor densities in the range \(\mathbf{\sim10^{10}}\ \mathrm{cm^{-3}}\). 
  Despite different origins, both yield localized confinement suitable for strong DP coupling and phonon transduction~\cite{Mei2024,Mei2022,san,math}.}
  \label{fig:deformation_comparison}
\end{figure}

The DPs associated with gate-defined QDs and with impurity-bound states in cryogenic Ge exhibit closely similar spatial envelopes (Figure~\ref{fig:deformation_comparison}). In both cases the confinement is strongly localized and, to leading order for device modeling, well captured by a Gaussian well. The gate-defined profile (solid blue) is set electrostatically by lithographic gates that reshape the local band edges, while the impurity-induced profile (dashed orange) arises intrinsically from the frozen-out fields of shallow donors / acceptors ( B, Al, Ga, P). Although their physical origins differ, the fitted depths and widths are comparable, implying that impurity-induced dipole states effectively emulate gate-defined dots in their ability to confine single carriers and support discrete energy levels. This equivalence makes impurity-bound dots a naturally occurring, reproducible, and low-noise platform for quantum sensing in high-purity Ge.

\subsection{Phononic crystal for slow-phonon guiding and thermal filtering}

A patterned, near–surface PnC layer (thickness $\sim\!\SI{1}{\micro m}$) creates acoustic bandgaps and slow–phonon bands that
(i) steer ballistic phonons toward the sensing region and
(ii) \emph{filter} thermally populated modes.
By tailoring the lattice, stop bands can suppress phonons near
$f_{\mathrm{th}}\!\simeq\!k_{B}T/h \approx \SI{83}{GHz}$ at \SI{4}{K},
while admitting the target GHz band where transduction is strongest~\cite{Joannopoulos2008}.
Embedding the localized states within (or adjacent to) a PnC cavity increases dwell time and strain participation at the dot, enhancing coupling without excessive metal participation losses.

\subsection{Phonon-to-charge transduction with RF-QPC}

A burst of athermal phonons modulates the local strain field and dot potential, producing a small displacement of the bound charge.
The nearby QPC senses this via the Ramo--Shockley mechanism as an induced charge
\begin{equation}
  Q_{\mathrm{ind}} \;\simeq\; Q\,\frac{\delta x}{d},
  \label{eq:qind}
\end{equation}
with carrier charge $Q$, effective displacement $\delta x$, and dot--QPC separation $d$.
Embedding the QPC in an RF reflectometry circuit converts $Q_{\mathrm{ind}}$ into a change in complex reflection coefficient, enabling high–bandwidth, sub–electron charge sensitivity compatible with nanosecond timing and GHz–band PSD selections.
Because the active Ge volume is free of through–bulk contacts and long drift paths, the architecture avoids common excess–noise and microphonic pickup mechanisms.

\begin{figure}[h]
  \centering
  \includegraphics[width=\linewidth]{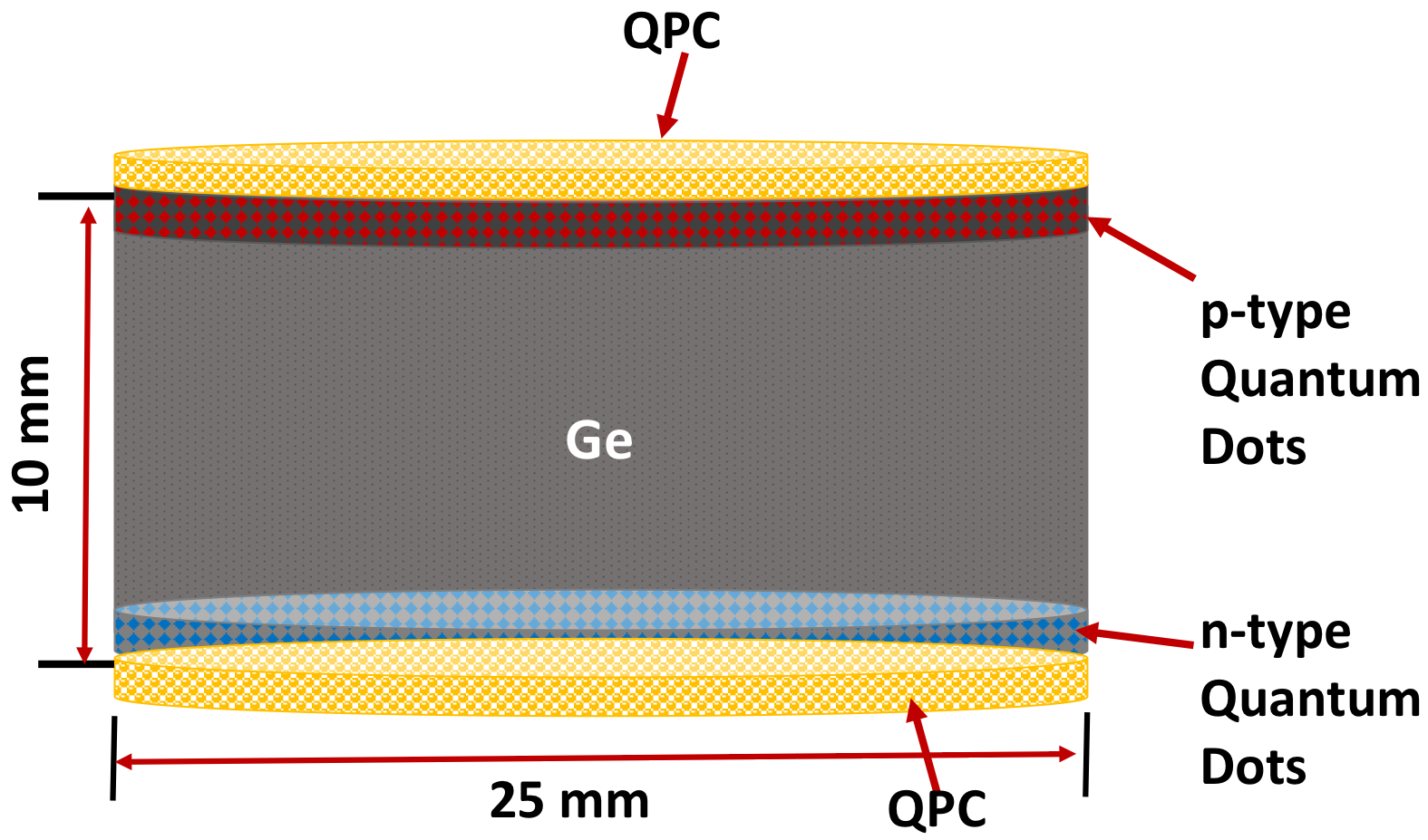}\\
  \includegraphics[width=\linewidth]{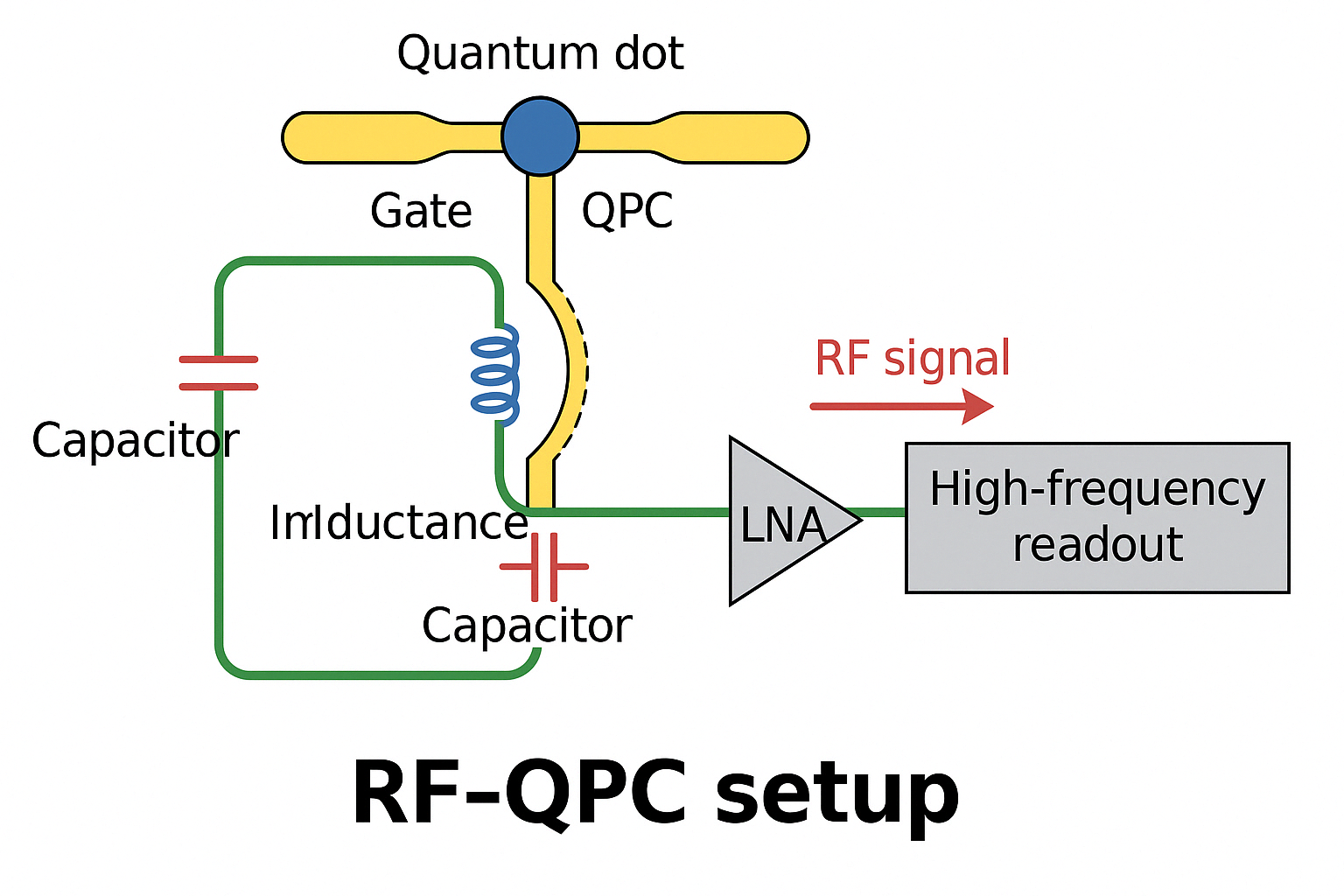}
  \caption{%
    \textbf{Top:} Schematic of the GeQuLEP detector architecture.
    A high-purity Ge crystal hosts spatially separated p-type (acceptor) and n-type (donor) dipole-bound regions, each co–located with a \(\sim\!\SI{1}{\micro m}\) PnC cavity that traps target GHz phonons and filters thermal backgrounds.
    \textbf{Bottom:} RF-QPC reflectometry.
    Phonon–induced charge motion modulates the QPC conductance and is read out as a change in the reflected RF carrier after cryogenic amplification and demodulation.}
  \label{fig:gequlep_readout}
\end{figure}

Figure~\ref{fig:gequlep_readout} shows the GeQuLEP detector and its RF-QPC readout. 
P-type and n-type dipole–bound regions are placed near opposite faces of a high-purity Ge crystal to provide charge sensitivity across the bulk. 
Each region is embedded in a near-surface PnC cavity that both confines and spectrally filters ballistic phonons, while lithographically aligned QPCs transduce the resulting phonon-induced charge motion. 
Together, PnC guiding and RF-QPC reflectometry convert low-energy phonon bursts into robust electrical signals, yielding a scalable, low-noise platform optimized for GHz-band phonon spectroscopy.

\subsection{Integrated design priorities}

The device layout follows three principles:
\begin{itemize}
  \item \textbf{Geometric return:} internal reflections and PnC steering to drive the geometrical phonon detection efficiency $\eta_{\mathrm{geom}}\!\to\!1$;
  \item \textbf{Propagation survival:} minimize parasitic absorption so phonon propagation survival probability $\eta_{\mathrm{prop}}=\exp(-n_d \sigma_{\mathrm{abs}} L)$ remains near unity for representative path length $L$;
  \item \textbf{Strong transduction:} maximize strain participation at the QD and QPC transconductance while keeping metal participation low and placing electrodes at strain nodes to preserve mechanical $Q$.
\end{itemize}

\section{Phonon Generation and Propagation in Ge}
\label{sec:phonon_physics}

Low energy particles, candidates for DM or neutrinos, deposit sub-eV to sub-keV recoil energies in high purity Ge. At cryogenic temperature ($T\!\sim\!\SI{4}{K}$), ionization is strongly suppressed by the Ge band gap ($\approx\SI{0.73}{eV}$) and the cost $\sim\!\SI{2.96}{eV}$ per electron-hole pair, so the \emph{dominant} energy channel enters quantized lattice vibrations (phonons)~\cite{Luke1989,Shutt1992}. Initial excitation produces \textbf{ primary phonons} predominantly in the longitudinal (LA) and transverse (TA) acoustic branches with typical energies of $\sim\!\SIrange{1}{10}{meV}$~\cite{Shank1983,Haller1974}. Because these recoil energies lie well below the $\sim\!\SIrange{30}{40}{meV}$ optical phonon threshold in Ge, optical emission is highly suppressed and long-wavelength acoustic modes are favored. Among them, LA modes often dominate early energy transport because of stronger deformation-potential coupling and a larger group velocity, which enhances their interaction with localized charge states used for transduction.

\paragraph{Microscopic picture and spectral scales}
To anchor scales, consider the standard 1D coupled–oscillator model (lattice constant $a$, atomic mass $m$, spring constant $k$)~\cite{Mei2025}:
\begin{equation}
  L \;=\; \sum_{i} \frac{1}{2}m\dot{u}_i^2 \;-\; \sum_{i} \frac{1}{2}k\,(u_i - u_{i-1})^2,
\end{equation}
with equations of motion
\begin{equation}
  m\ddot{u}_i \;=\; -k(u_i - u_{i-1}) + k(u_{i+1} - u_i).
\end{equation}
A traveling–wave ansatz $u_i(t)=A\,e^{i(\omega t - q u_i)}$ yields the acoustic dispersion
\begin{equation}
  \omega(q) \;=\; 2\sqrt{\frac{k}{m}}\;\sin\!\left(\frac{qa}{2}\right), 
  \qquad
  \omega_{\max} \;=\; 2\sqrt{\frac{k}{m}},
\end{equation}
where $\omega$ is the angular
frequency and $q$ is the wavevector. The maximum vibrational frequency is attained at the
Brillouin zone edge ($q$ = $\pi$/a).

Using $m\!\approx\!1.206\times10^{-25}\ \mathrm{kg}$ for Ge and an effective $k\!\approx\!\SI{3.86}{N/m}$ (consistent with Ge acoustic branches) gives
$\omega_{\max}\!\approx\!1.13\times10^{13}\ \mathrm{rad/s}$, i.e.
\begin{equation}
  f_{\max} \;=\; \frac{\omega_{\max}}{2\pi} \;\approx\; \SI{1.80}{THz}, 
  \end{equation}
  \begin{equation}
      E_{\max} \;=\; h f_{\max} \;\approx\; \SI{7.45}{meV},
  \end{equation}
consistent with measured acoustic phonon energies in Ge~\cite{YuCardona2010,Haller1974,Agnese2018}.
\emph{The full DM–induced phonon emission rate, branch decomposition, and normalization used in our projections are derived in Mei et al.~\cite{Mei2018}} (including DM–electron and CE$\nu$NS–like nuclear channels).

\paragraph{Anharmonic down–conversion and ballistic regime}
Immediately after production, high–frequency primary phonons relax through three–phonon (\mbox{$\omega_0\!\to\!\omega_1+\omega_2$}) processes that conserve energy and crystal momentum,
\begin{equation}
  \omega_0=\omega_1+\omega_2, \qquad \vec{q}_0=\vec{q}_1+\vec{q}_2,
\end{equation}
cascading energy into a burst of lower–frequency \emph{athermal} acoustic modes~\cite{Haller1974,Irwin1995}.
As the spectrum softens below $\sim\!\SI{1}{meV}$, scattering rates drop sharply; at $T\!\lesssim\!\SI{10}{K}$, LA/TA modes in ultra–pure Ge propagate quasi–ballistically over centimeter scales, preserving the temporal and spatial imprint of the originating recoil~\cite{Irwin1995,Haller1974}.
Crystal anisotropy produces pronounced \emph{phonon focusing}, channeling energy along high–symmetry directions~\cite{Liu1999}.
Residual elastic scattering from isotopes and dilute defects increases steeply with frequency, so low–GHz modes dominate long–range transport~\cite{Tamura1991,Perrin2006}.

\begin{figure}[h]
  \centering
  \includegraphics[width=0.48\textwidth]{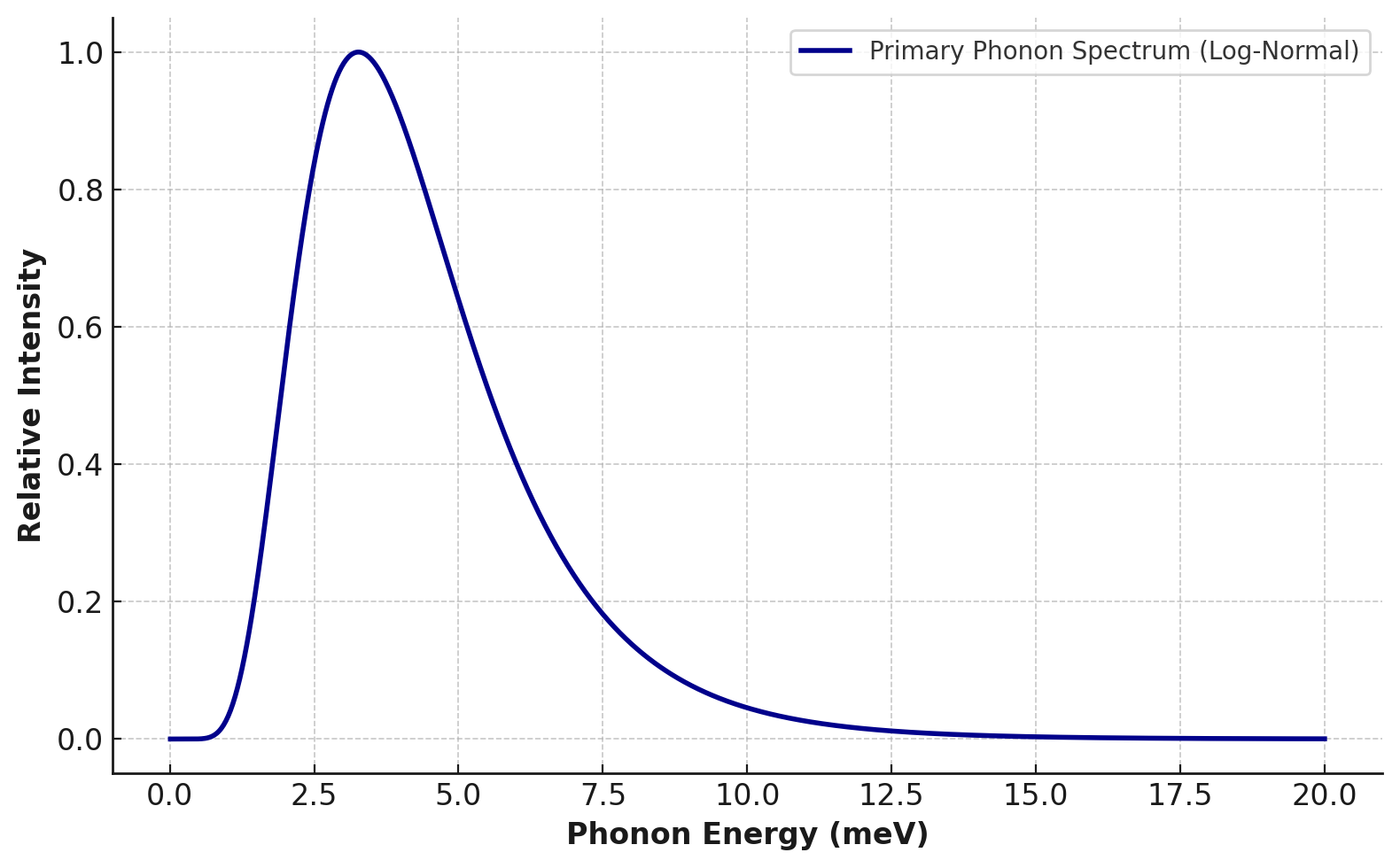}\hfill
  \includegraphics[width=0.48\textwidth]{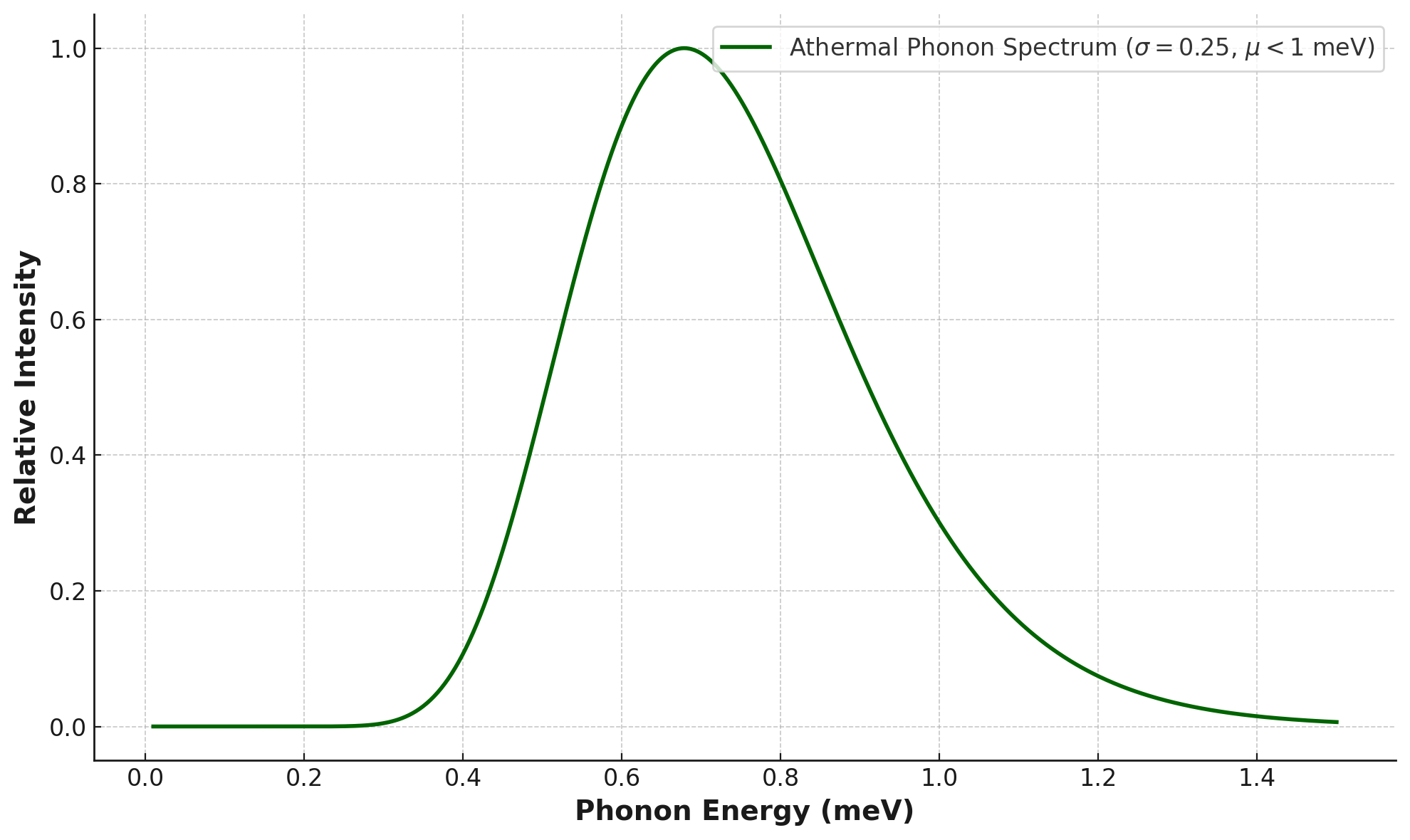}
  \caption{%
  \textbf{Top:} Schematic energy spectrum of primary (LA/TA) phonons produced by a localized recoil in Ge. The initial primary-phonon energy distribution is represented by a log-normal function parameterized by a median energy $E_0$ and logarithmic width $\sigma_{\ln E}$; this empirical fit seeds the cascade model and aids visualization.
  \textbf{Bottom:} Athermal spectrum after anharmonic down–conversion, with weight shifted to lower–frequency modes that propagate ballistically at cryogenic temperature.}
  \label{fig:phonon_evolution}
\end{figure}

Figure~\ref{fig:phonon_evolution} traces the evolution in high-purity Ge from a narrow, high-frequency \emph{primary} spectrum to a broader, lower-energy \emph{athermal} distribution. At few-kelvin temperatures these long-mean-free-path acoustic modes propagate quasi-ballistically over centimeter scales, preserving nanosecond timing and spatial information about the recoil. GeQuLEP exploits this behavior: a near-surface PnC cavity steers the returning flux into the sensing region; localized gate-defined or impurity-bound states absorb the strain via deformation-potential coupling; and an adjacent RF-QPC operated in reflectometry converts the phonon-induced charge motion into a measurable change in the complex reflection coefficient. This phonon-to-charge chain underpins the sub-eV threshold and the GHz-band timing/PSD selections used throughout this work.

\paragraph{Implications for the GeQuLEP design}
The GeQuLEP architecture is optimized to harvest this athermal burst: a near–surface PnC cavity slows and guides returning acoustic energy into the sensing region, while the localized charge state (gate–defined or impurity–bound) provides strong deformation–potential coupling for phonon–to–charge transduction.
This section summarizes the physics inputs used in our efficiency and sensitivity calculations; the \emph{detailed} derivation of the DM–induced phonon emission, phase–space limits, and branch–resolved spectra. The next section analyzes boundary reflectivity at the Ge–vacuum interface and its role in phonon confinement and waveguiding within the device.

\section{Acoustic reflectivity and guiding at the Ge--vacuum boundary}
\label{sec:reflectivity}

A free Ge surface (Ge--vacuum) enforces a \emph{traction-free} boundary condition, $\boldsymbol{\sigma}\cdot\hat{\boldsymbol{n}}=0$, because vacuum cannot support acoustic stress. In the plane–wave (impedance) picture this is equivalent to a nearly perfect acoustic mirror. The specific acoustic impedance of a mode is
\begin{equation}
  Z = \rho\,v,
  \label{eq:impedance}
\end{equation}
with mass density $\rho$ and mode velocity $v$. For LA and TA modes in Ge at cryogenic temperature, representative values are
\[
  Z^{\mathrm{LA}}_{\mathrm{Ge}} \simeq \rho_{\mathrm{Ge}} v_L \approx \SI{2.9e7}{kg\,m^{-2}\,s^{-1}}, 
  \]\\
  \[
  Z^{\mathrm{TA}}_{\mathrm{Ge}} \simeq \rho_{\mathrm{Ge}} v_T \approx \SI{1.9e7}{kg\,m^{-2}\,s^{-1}},
\] \\
using $\rho_{\mathrm{Ge}}\!\approx\!\SI{5323}{kg\,m^{-3}}$, $v_L\!\approx\!\SI{5.4}{km\,s^{-1}}$, and $v_T\!\approx\!\SI{3.5}{km\,s^{-1}}$. Vacuum has $Z_{\mathrm{vac}}=0$.

\paragraph{Normal incidence}
For a longitudinal wave normally incident on an interface between media of impedances $Z_1$ and $Z_2$, the \emph{power} reflectivity is
\begin{equation}
  R \;=\; \left(\frac{Z_2 - Z_1}{Z_2 + Z_1}\right)^2.
  \label{eq:R_normal}
\end{equation}
At the Ge--vacuum boundary, $Z_2\!=\!0$, so $R\!=\!1$ and the \emph{amplitude} reflection coefficient $r\!=\!-1$ (a $\pi$ phase flip). This ideal limit is consistent with cryogenic phonon–detector experience on optically polished Ge, which shows negligible energy loss across free surfaces~\cite{Mazin2002,Haller1974}.

\paragraph{Oblique incidence, mode conversion, and surface waves}
For elastic solids, an obliquely incident LA (or TA) wave can reflect as a mixture of LA and TA waves, determined by the traction–free boundary condition (Zoeppritz–like relations in elasticity)~\cite{Cleland2003}. Because the “second medium’’ is vacuum, Snell’s law admits no propagating transmitted solution; hence \emph{total internal reflection} holds for all incident angles. A fraction of the energy can couple into a \emph{Rayleigh} surface mode localized near the free surface; the partition depends on incidence angle and polarization~\cite{Cleland2003}. In GeQuLEP, the near–surface PnC is designed to intercept returning bulk modes and to engineer the surface dispersion so that lossy channels are minimized.

\paragraph{Specularity vs.\ diffuse scattering (surface quality)}
Departures from perfect reflection are governed primarily by surface quality at wavelengths of interest. For a surface with rms roughness $\eta$ and an incident acoustic wavelength $\lambda$, Ziman’s specularity parameter
\begin{equation}
  p(\lambda,\theta) \;\approx\; \exp\!\left[-\,\frac{16\pi^{2}\eta^{2}\cos^{2}\theta}{\lambda^{2}}\right]
  \label{eq:ziman}
\end{equation}
estimates the probability of \emph{specular} reflection at incidence angle $\theta$~\cite{Klitsner1988}. In the \SIrange{10}{100}{GHz} band relevant here, $\lambda\!=\!v/f$ lies in the \SIrange{50}{500}{nm} range for Ge; maintaining $\eta\!\lesssim\!\SI{1}{nm}$ yields $p$ close to unity over most angles. This motivates optical–grade polishing and chemical passivation to suppress diffuse scattering and conversion into lossy surface or amorphous modes~\cite{Klitsner1988,Haller1974}.

\paragraph{Thin films, coatings, and Fabry--Pérot effects}
Thin layers (e.g., native oxides, metal electrodes) modify boundary conditions and can introduce interference akin to Fabry--Pérot resonances. For a film of thickness $t$ bounded by high–reflectivity interfaces, constructive conditions
\begin{equation}
  2\,t\,\cos\theta \;=\; m\,\lambda, \qquad m\in\mathbb{Z}^{+},
  \label{eq:FP}
\end{equation}
produce frequency–dependent standing waves that modulate local energy density and effective reflectivity~\cite{SwartzPohl1989}. While such resonances can be exploited deliberately (e.g., to enhance dwell time near the sensor), uncontrolled films or rough overlayers can broaden resonances and increase loss. In GeQuLEP, we therefore (i) minimize lossy overlayer thickness, (ii) place metal at displacement or strain nodes to reduce participation, and (iii) rely on the PnC cavity to set the dominant spectral response instead of uncontrolled multilayer optics.

\paragraph{Guiding and geometric return}
Because $R\!\approx\!1$ and inelastic bulk scattering is strongly suppressed at cryogenic temperature, ballistic LA/TA modes undergo many specular bounces, repeatedly returning energy toward the surface. The PnC adjacent to the sensing region provides (a) angular acceptance matched to phonon focusing patterns, (b) band-edge slowing to increase dwell time and strain participation, and (c) stop bands that attenuate thermally populated modes (notably near $f_{\mathrm{th}}\simeq k_BT/h\approx\SI{83}{GHz}$ at \SI{4}{K})~\cite{SwartzPohl1989,Cleland2003}. This combination of high reflectivity and engineered guiding underpins the high geometric collection factor used in our efficiency model.

\paragraph{Design implications for GeQuLEP}
To preserve near-ideal reflectivity and coherence we adopt:
(i) sub-nanometer rms surface roughness with optical–grade polishing and in situ passivation,
(ii) minimal and well–placed metal to avoid acoustic loading,
(iii) thin, well–controlled dielectric/oxide layers, and
(iv) a PnC cavity that determines the usable band and suppresses thermal backgrounds.
These measures maximize phonon \emph{geometric return} and \emph{propagation survival}, enabling efficient delivery of athermal flux to the localized charge sensor and, ultimately, sub–eV thresholds.


\section{Phonon collection model and primary-phonon detection}
\label{sec:collection}

We model the probability that a \emph{single} athermal phonon produced in the bulk is ultimately sensed as induced charge in the QPC as a product of (i) geometric delivery to the sensing region, (ii) survival along the path, and (iii) local capture in (or adjacent to) the quantum-well (QW) / PnC cavity:
\begin{equation}
  \eta_{\mathrm{coll}}(\omega,\Omega) \;=\; 
  \underbrace{\eta_{\mathrm{geom}}(\Omega)}_{\text{guiding \& return}}
  \;\times\;
  \underbrace{\eta_{\mathrm{prop}}(\omega)}_{\text{bulk survival}}
  \;\times\;
  \underbrace{\eta_{\mathrm{ath}}(\omega)}_{\text{local capture}},
  \label{eq:eta_coll_factorized}
\end{equation}
where $\omega$ and $\Omega$ denote frequency and direction. Event–level detection will then fold in the \emph{primary} multiplicity $M$ from anharmonic down–conversion (Sec.~\ref{sec:phonon_physics}) and the number of primaries per recoil.

\subsection{Geometric collection: guiding, return, and acceptance}
Primary acoustic phonons are emitted broadly in angle from a localized recoil. A naive single–pass solid–angle estimate,
$\eta_{\mathrm{geom}}\!\approx\!\Omega_{\mathrm{acc}}/4\pi$,
severely underestimates delivery because the Ge--vacuum interface is an almost perfect acoustic mirror (Sec.~\ref{sec:reflectivity}). Multiple specular bounces, combined with phonon focusing in crystalline Ge, iteratively redirect energy back toward the near–surface sensing region. A convenient way to encode this is as a hitting–probability with $N$ effective encounters,
\begin{equation}
  \eta_{\mathrm{geom}} \;\simeq\; 1 - \left(1 - f_{\mathrm{acc}}\;p\right)^{N},
  \label{eq:eta_geom_multihit}
\end{equation}
where $f_{\mathrm{acc}}$ is the fractional angular/areal acceptance of the PnC/QW region for a returning flux and $p$ is the specularity parameter of the surface (Eq.~\ref{eq:ziman}). With optical–grade polishing ($p\!\gtrsim\!0.95$ in the \SIrange{10}{100}{GHz} band) and tens–to–hundreds of effective encounters before bulk loss or thermalization, $\eta_{\mathrm{geom}}$ approaches unity. The PnC further boosts $f_{\mathrm{acc}}$ by steering and band–edge slowing, so in optimized layouts we adopt $\eta_{\mathrm{geom}}\!\to\!(0.9$–$1.0)$ as a realistic range.

\subsection{Propagation survival in bulk Ge}
The dominant loss along the path is parasitic absorption on unintended dipole or defect states in the bulk (elastic isotope scattering primarily redistributes direction but is weak at low $\omega$). Modeling absorption with Beer–Lambert attenuation gives~\cite{Mei2025}
\begin{equation}
  \eta_{\mathrm{prop}}(\omega) \;=\; 
  \exp\!\Big[-\,n_d\,\sigma_{\mathrm{abs}}(\omega)\,L_{\mathrm{eff}}(\omega)\Big],
  \label{eq:eta_prop}
\end{equation}
where $n_d$ is the density of absorbers, $\sigma_{\mathrm{abs}}$ the deformation–potential–mediated absorption cross section, and $L_{\mathrm{eff}}$ an effective path length that includes multiple bounces (equivalently, $L_{\mathrm{eff}}=v_g\,t_{\mathrm{dwell}}$ with group velocity $v_g$ and dwell time set by geometry and surface specularity).  Applying Fermi’s golden rule to LA phonons interacting with a localized two–level center~\cite{Ziman1972,Mahan2000,Cleland2003}, and normalizing by the incident phonon flux, yields the expression below; a detailed derivation is provided in Mei \emph{et~al.}~\cite{Mei2025}.

\begin{equation}
  \sigma_{\mathrm{abs}}(\omega) \;\simeq\; 
  \frac{D^2}{\rho\,v_g^3\,\hbar}\;\times\;F(\omega),
  \label{eq:sigma_abs}
\end{equation}
where $D$ is the deformation–potential constant, $\rho$ the mass density, and $F(\omega)$ captures line shape and level–structure factors (unit–order away from resonances). The strong $v_g^{-3}$ dependence implies that slower phonons are far more susceptible to absorption. For representative cryogenic Ge values ($D_e\!\approx\!\SI{13.5}{eV}$ for electrons, $D_h\!\approx\!\SI{4.5}{eV}$ for holes) and bulk $v_g$, one finds
$\sigma_{\mathrm{abs}}\!\sim\!5\times10^{-13}\ \mathrm{cm^2}$ (electrons) and $\sim\!6\times10^{-14}\ \mathrm{cm^2}$ (holes)~\cite{Mahan2000,SwartzPohl1989,Klitsner1988}.
Adopting $n_d\!\sim\!10^{10}\ \mathrm{cm^{-3}}$ and $L_{\mathrm{eff}}\!\sim\!10\ \mathrm{cm}$ gives
$\eta_{\mathrm{prop}}\!\approx\!0.95$, i.e., only a few–percent loss even in thick devices; cleaner crystals and shorter $L_{\mathrm{eff}}$ drive this closer to unity.

\subsection{Local capture (athermal) in the QW/PnC region}
\begin{figure}[h]
    \centering
    \includegraphics[width=0.85\linewidth]{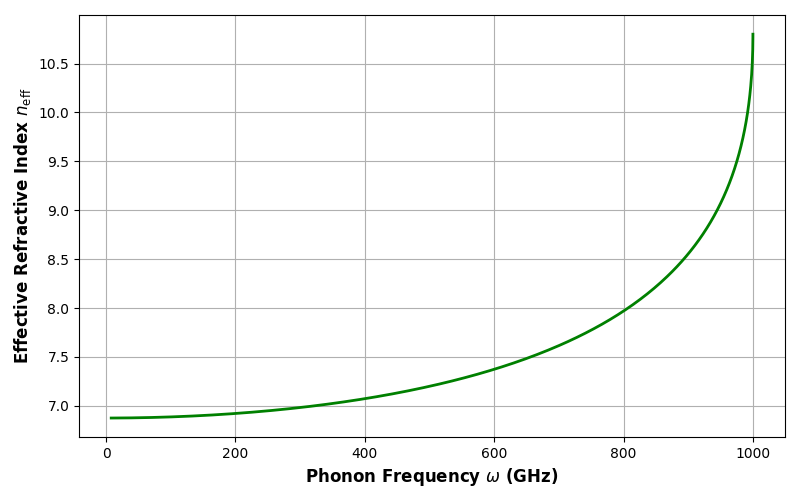}
    \caption{Effective refractive index $n_{\mathrm{eff}}$ as a function of phonon frequency $\omega$ based on a synthetic dispersion relation in a SiGe PnC cavity.}
    \label{fig:neff}
\end{figure}

\begin{figure}[h]
    \centering
    \includegraphics[width=0.85\linewidth]{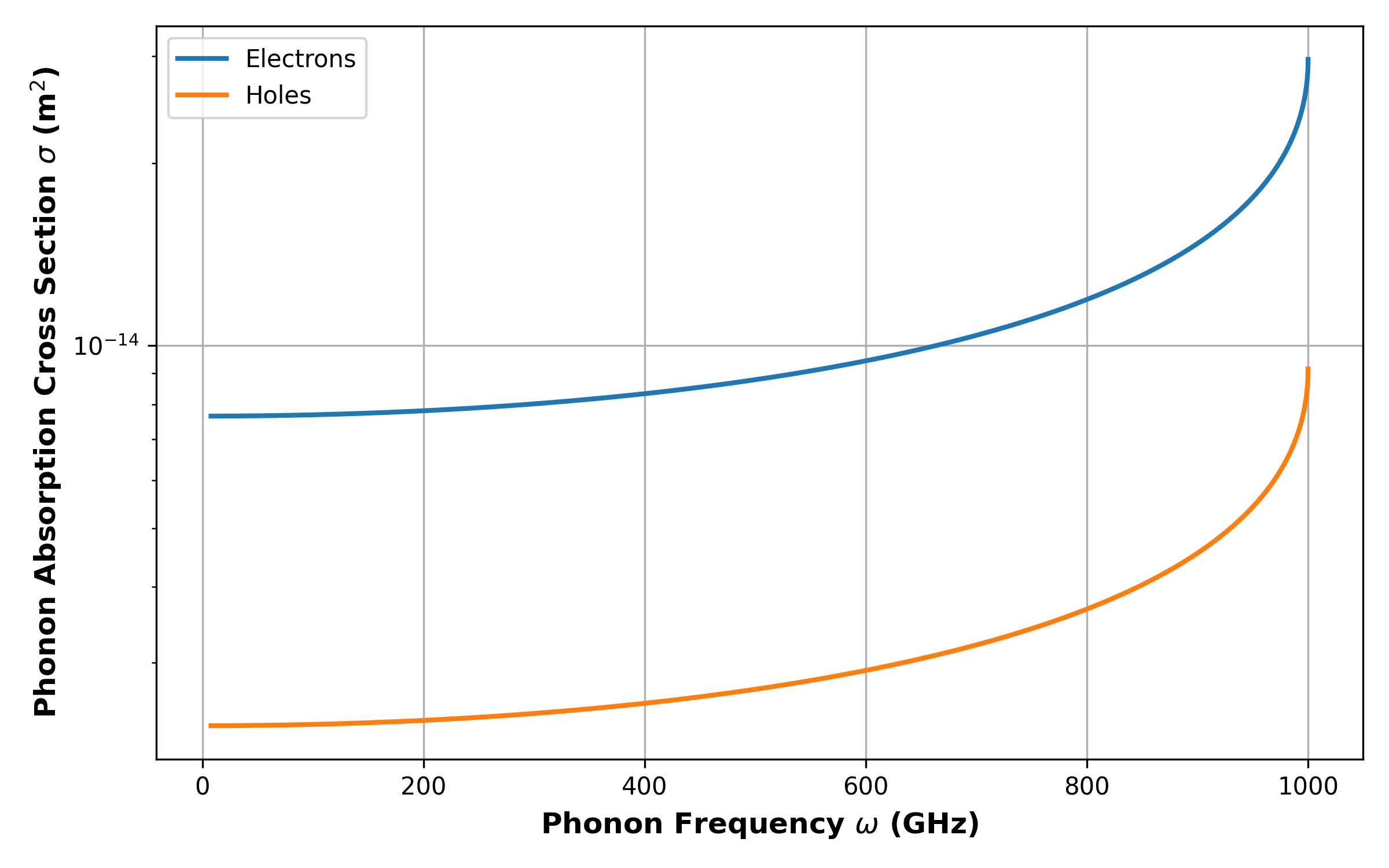}
    \caption{Phonon absorption cross section $\sigma$ as a function of phonon frequency in the quantum well region. The enhanced absorption at higher frequencies corresponds to the regime where phonons are strongly confined and slowed, increasing the likelihood of interaction with dipole states.}
    \label{fig:sigma_qw}
\end{figure}

The per–phonon capture probability in (or near) the QW is
\begin{equation}
  \eta_{\mathrm{ath}}(\omega) \;=\; 1 - 
  \exp\!\Big[-\,n_{\ell}\,\sigma_{\mathrm{abs}}^{(\mathrm{loc})}(\omega)\,\ell_{\mathrm{int}}\Big],
  \label{eq:eta_ath}
\end{equation}
where $n_{\ell}$ is the local density of active dipole states, $\ell_{\mathrm{int}}$ the interaction length across the sensing region, and $\sigma_{\mathrm{abs}}^{(\mathrm{loc})}$ is the \emph{local} cross section. The PnC cavity modifies the dispersion near band edges, reducing $v_g$ and thereby enhancing $\sigma_{\mathrm{abs}}$ via Eq.~\eqref{eq:sigma_abs}. This “slow–phonon’’ enhancement is conveniently expressed with an effective index $n_{\mathrm{eff}}(\omega)=v_{\mathrm{bulk}}/v_g(\omega)$, so that
$\sigma_{\mathrm{abs}}^{(\mathrm{loc})}\!\propto\!n_{\mathrm{eff}}^3$.
As shown in Figure~\ref{fig:neff}, $n_{\mathrm{eff}}$ grows rapidly where the PnC band flattens, and the corresponding cross section (Figure~\ref{fig:sigma_qw}) increases sharply, driving $\eta_{\mathrm{ath}}$ toward $\mathcal{O}(0.1$–$0.3)$ for micron–scale $\ell_{\mathrm{int}}$ at practical $n_{\ell}$.

\subsection{Multiplicity-assisted \emph{primary} detection}
A high–frequency \emph{primary} phonon undergoes anharmonic decay into $M$ lower–frequency daughters (Sec.~\ref{sec:phonon_physics}). If each daughter is captured with probability $\eta_{\mathrm{ath}}$ (averaged over the signal band), the probability to detect \emph{at least one} daughter from a given primary is
\begin{equation}
  \eta_{\mathrm{primary}} \;=\; 1 - \big(1 - \eta_{\mathrm{ath}}\big)^M.
  \label{eq:eta_primary}
\end{equation}
For $\eta_{\mathrm{ath}}\!\sim\!0.10$ and $M\!\approx\!16$, one obtains $\eta_{\mathrm{primary}}\!\approx\!0.81$, clarifying how one may \emph{count} “$>\!1$ athermal phonon per primary’’ without violating energy conservation: multiplicity arises before capture, while the energy efficiency remains $\le\!100\%$.

\subsection{Spectral averaging and event-level detection}
The collected athermal spectrum $S(\omega,\Omega)$ (normalized to unity) weights Eqs.~\eqref{eq:eta_coll_factorized}–\eqref{eq:eta_primary}:
\begin{equation}
  \langle \eta_{\mathrm{coll}} \rangle \;=\; 
  \iint S(\omega,\Omega)\,\eta_{\mathrm{coll}}(\omega,\Omega)\,d\omega\,d\Omega,
  \end{equation}
  \begin{equation}
  \langle \eta_{\mathrm{primary}} \rangle \;=\;
  \int S(\omega)\,\Big[1-(1-\eta_{\mathrm{ath}}(\omega))^{M(\omega)}\Big]\,d\omega.
\end{equation}
For a recoil producing $N_{\mathrm{prim}}$ primaries, the probability that \emph{none} lead to a detected daughter after geometric delivery and survival is
$P_0=\big[1-\langle \eta_{\mathrm{geom}}\eta_{\mathrm{prop}}\rangle\,\langle \eta_{\mathrm{primary}}\rangle\big]^{N_{\mathrm{prim}}}$,
so the event–level detection probability is
\begin{equation}
  P_{\mathrm{det}} \;=\; 1 - P_0 
  \;=\; 1 - \Big(1-\langle \eta_{\mathrm{geom}}\eta_{\mathrm{prop}}\rangle\,\langle \eta_{\mathrm{primary}}\rangle\Big)^{N_{\mathrm{prim}}}.
  \label{eq:P_event}
\end{equation}
In the low–yield limit ($N_{\mathrm{prim}}\langle\cdot\rangle\ll1$), $P_{\mathrm{det}}\!\approx\!N_{\mathrm{prim}}\langle \eta_{\mathrm{geom}}\eta_{\mathrm{prop}}\rangle\,\langle \eta_{\mathrm{primary}}\rangle$.

\paragraph{Representative numbers}
Using $n_d\!=\!10^{10}\ \mathrm{cm^{-3}}$, $\sigma_{\mathrm{abs}}\!=\!5\times10^{-13}\ \mathrm{cm^2}$ (electrons), and $L_{\mathrm{eff}}\!=\!10\ \mathrm{cm}$ gives $\eta_{\mathrm{prop}}\!\approx\!0.95$.
With polished surfaces ($p\!\gtrsim\!0.95$) and PnC steering, $\eta_{\mathrm{geom}}\!\to\!0.9$–$1.0$.
Taking $\eta_{\mathrm{ath}}\!\sim\!0.10$ (micron–scale QW with slow–phonon enhancement) and $M\!=\!16$ yields $\eta_{\mathrm{primary}}\!\approx\!0.81$.
Combined, a realistic \emph{per–primary} detection factor
$\langle \eta_{\mathrm{geom}}\eta_{\mathrm{prop}}\rangle\,\langle \eta_{\mathrm{primary}}\rangle$
in the $0.7$–$0.8$ range is attainable for optimized devices, with headroom for improvement via higher $n_{\mathrm{eff}}$, larger QW participation, and reduced $n_d$.

\paragraph{Design levers}
(i) \emph{Geometry:} enlarge PnC acceptance and maintain high specularity to push $\eta_{\mathrm{geom}}\!\to\!1$.
(ii) \emph{Bulk purity:} reduce $n_d$ to raise $\eta_{\mathrm{prop}}$ toward unity.
(iii) \emph{Slow–phonon engineering:} use band–edge flattening to increase $n_{\mathrm{eff}}$ and thus $\sigma_{\mathrm{abs}}^{(\mathrm{loc})}$, boosting $\eta_{\mathrm{ath}}$.
(iv) \emph{Localization:} maximize strain participation at the dot and interaction length $\ell_{\mathrm{int}}$ without adding lossy metal participation.


\section{QPC readout and induced-charge response}
\label{sec:readout}

Phonon-induced modulation of the local strain field shifts the energy landscape of a localized carrier (gate-defined or impurity-bound), producing a small displacement $\delta x$ of the bound charge. The adjacent radio-frequency quantum point contact (RF-QPC), operated in reflectometry, senses this motion as an \emph{induced charge} via the Ramo--Shockley theorem. Within a parallel–plate weighting–field approximation (valid for $\delta x \ll d$), Eq.~\ref{eq:qind} --
  $Q_{\mathrm{ind}} \simeq Q\,\frac{\delta x}{d}$ applies,
where $Q$ is the carrier charge, $\delta x$ is the displacement along the weighting field, and $d$ is the effective dot–QPC separation.
 Device-scale calculations (Sec.~\ref{sec:collection}) and the driven-oscillator model below indicate a pronounced response band in the \SI{10}{-}\SI{30}{GHz} region associated with dipole-bound oscillations, reaching $Q_{\mathrm{ind}}\!\sim\!\mathcal{O}(10^{-2})\,e$ for $d\!\sim\!\SI{1}{\micro m}$ and remaining above $10^{-3}\,e$ even near the $\sim\!\SI{125}{GHz}$ ballistic band—well within modern RF-QPC sensitivity~\cite{Field1993,Schoelkopf1998,GonzalezZalba2015}.

\subsection{Charge--phonon coupling and slow-phonon enhancement}
\begin{figure}[h]
  \centering
  \includegraphics[width=0.45\textwidth]{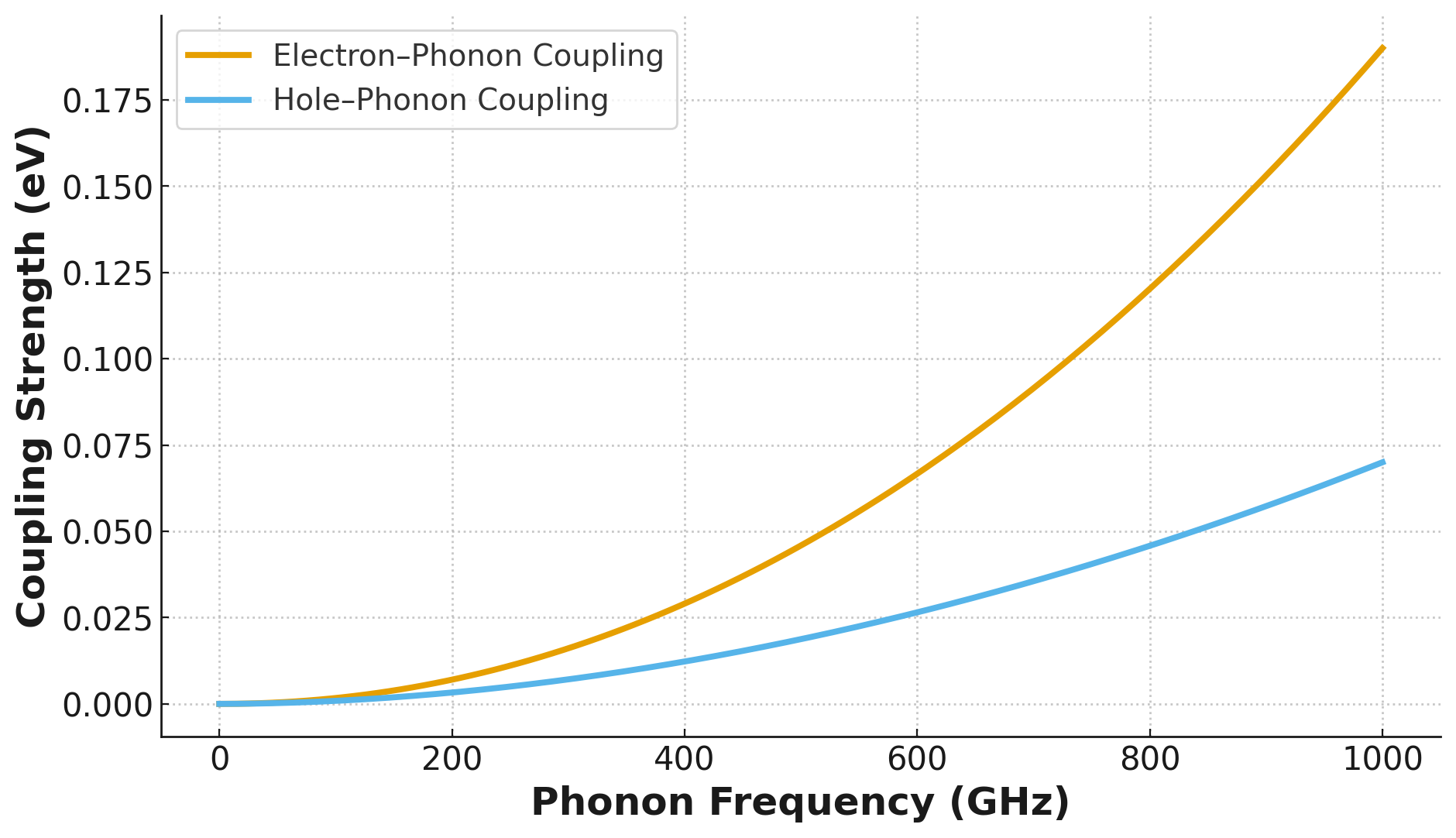}
  \caption{Charge-phonon coupling strength $g(\omega)$ in Ge versus frequency, using Eq.~\eqref{eq:gomega} and PnC-enhanced mode volumes. The slow-phonon regime boosts $g$ via reduced $v_{\mathrm{ph}}$ and $V_{\mathrm{mode}}$.}
  \label{fig:phonon_coupling}
\end{figure}

Phonons couple to localized carriers through the DP: local dilational strain $\mathrm{Tr}\,(\varepsilon$) shifts the dot levels by $\Delta E_{\mathrm{DP}}=D\,\mathrm{Tr}\,(\varepsilon$) (Sec.~\ref{sec:concept_design}). Quantizing the acoustic mode in a volume $V_{\mathrm{mode}}$ yields a single-phonon coupling strength~\cite{YuCardona2010,Mahan2000, Mei2025}
\begin{equation}
  g(\omega) \;=\; D\,\frac{\omega}{v_{\mathrm{ph}}}\,
  \sqrt{\frac{\hbar}{2\rho V_{\mathrm{mode}}\omega}}
  \;=\; D\,\sqrt{\frac{\hbar\,\omega}{2\rho V_{\mathrm{mode}}\,v_{\mathrm{ph}}^{2}}},
  \label{eq:gomega}
\end{equation}
with mass density $\rho$, phase velocity $v_{\mathrm{ph}}$, and angular frequency $\omega=2\pi f$. In a PnC cavity, band-edge flattening slows the mode [$v_{\mathrm{ph}}\!\to\!v_{\mathrm{ph}}/n_{\mathrm{eff}}(\omega)$] and compresses the modal volume~\cite{Joannopoulos2008,Laude2015,Khelif2016}. A useful scaling is
\begin{equation}
  V_{\mathrm{mode}} \;\sim\; \Big(\frac{\lambda}{n_{\mathrm{eff}}}\Big)^{3}, 
  \qquad \lambda=\frac{v_{\mathrm{ph}}}{f},
\end{equation}
so that slow-phonon engineering boosts $g(\omega)$ through both smaller $v_{\mathrm{ph}}$ and $V_{\mathrm{mode}}$. 
As an illustration, at $f=\SI{30}{GHz}$, $\lambda\!\approx\!\SI{180}{nm}$ in bulk Ge; with $n_{\mathrm{eff}}=6.8$ (high-index-contrast, slow-phonon PnC), one finds 
$V_{\mathrm{mode}}\!\sim\!(\SI{180}{nm}/6.8)^3\!\approx\!1.9\times10^{-5}\ \mu\mathrm{m}^3$, consistent with high-$Q$ phononic resonators~\cite{Mohammadi2009,Hatanaka2014,Arrangoiz2016}. The net scaling $g(\omega)\!\propto\!\sqrt{\omega}$ implies stronger coupling at higher frequency (Figure~\ref{fig:phonon_coupling}).

\subsection{Single-phonon lattice displacement and driven bound-carrier response}
\begin{figure}[h]
  \centering
  \includegraphics[width=0.45\textwidth]{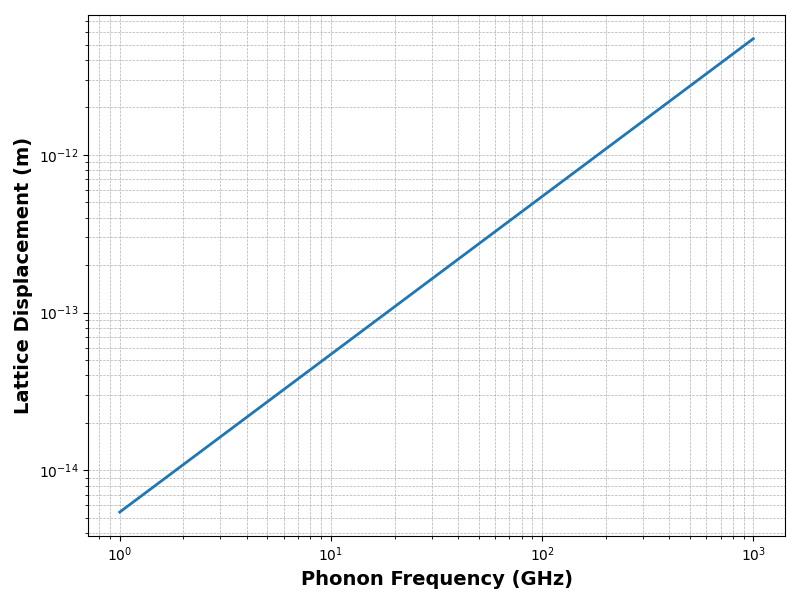}
  \caption{Single-phonon lattice displacement $u_0(\omega)$ from Eq.~\eqref{eq:u0}. Displacement decreases with frequency, but slow-phonon PnC confinement reduces $V_{\mathrm{mode}}$ and increases $u_0$ at a fixed $\omega$.}
  \label{fig:lattice_displacement}
\end{figure}

The rms lattice displacement for a single acoustic phonon in volume $V_{\mathrm{mode}}$ is~\cite{Mahan2000,Kittel2005}
\begin{equation}
  u_0(\omega) \;=\; \sqrt{\frac{\hbar}{2\rho V_{\mathrm{mode}}\,\omega}} \;\propto\; \omega^{-1/2}.
  \label{eq:u0}
\end{equation}
In our frequency range, $u_0\!\sim\!10^{-14}$–$10^{-13}\ \mathrm{m}$ for representative cryogenic Ge parameters and PnC mode volumes (Figure~\ref{fig:lattice_displacement}). This strain drives the bound carrier as a (weakly damped) driven oscillator~\cite{Mei2025},
\begin{equation}
  m\ddot{x} + m\gamma\dot{x} + kx \;=\; F_0\cos\omega t,
  \qquad \omega_0=\sqrt{k/m},
\end{equation}
with effective mass $m$, spring constant $k$ (set by the local confinement), and damping $\gamma$. For DP driving, a plane-wave estimate gives $F_0 \simeq D\,q^2\,u_0$ with $q=2\pi/\lambda$~\cite{Mahan2000}. The steady-state amplitude is
\begin{equation}
  \delta x(\omega) \;=\; \frac{D\,q^2\,u_0/m}{\sqrt{(\omega_0^2-\omega^2)^2+(\gamma\omega)^2}}
  \;\xrightarrow{\ \gamma\to0\ }\; \frac{D\,q^2\,u_0/m}{|\omega_0^2-\omega^2|}.
  \label{eq:dx}
\end{equation}
Resonant enhancement occurs when $\omega\!\approx\!\omega_0$; device design (dot size, material, and local strain) can place $\omega_0$ within the \SI{10}{-}\SI{30}{GHz} band, producing the strong response observed in Sec.~\ref{sec:collection} (Figure~\ref{fig:charge_displacement}). A more detailed derivation is provided in Mei \emph{et~al.}~\cite{Mei2025}.

\subsection{From displacement to measured signal: induced charge and RF readout}
\begin{figure}[h]
  \centering
  \includegraphics[width=0.45\textwidth]{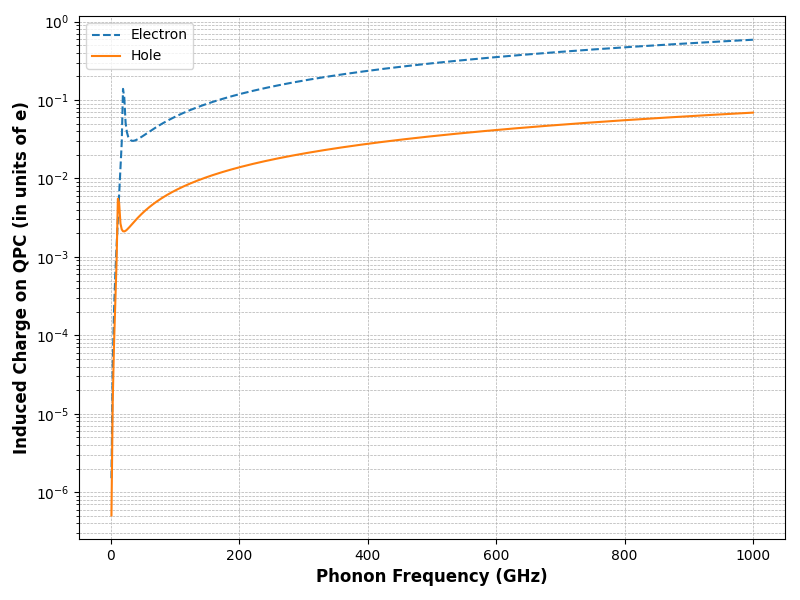}
  \caption{Modeled induced charge $Q_{\mathrm{ind}}(\omega)$ using Eq.~\eqref{eq:qind} with $d=\SI{1}{\micro m}$. A resonance band around \SI{10}{-}\SI{30}{GHz} maximizes response; near the ballistic band ($\sim\!\SI{125}{GHz}$) the signal remains $>\!10^{-3}\,e$.}
  \label{fig:induced_charge}
\end{figure}

Combining Eqs.~\eqref{eq:qind} and \eqref{eq:dx} gives the frequency-dependent induced charge on the QPC. Figure~\ref{fig:induced_charge} shows $Q_{\mathrm{ind}}(\omega)$ for electrons and holes at $d=\SI{1}{\micro m}$, with a pronounced \SI{10}{-}\SI{30}{GHz} resonance and a high-frequency tail that remains $>\!10^{-3}\,e$ near $\sim\!\SI{125}{GHz}$. In reflectometry, the QPC conductance $G(v_g)$ is embedded in an $LC$ tank; small $\Delta G$ modulates the complex reflection coefficient $r(\omega_{\mathrm{RF}})$, which is amplified cryogenically and demodulated at room temperature~\cite{Field1993,Schoelkopf1998}. State-of-the-art RF--QPCs resolve charge steps well below $10^{-3}\,e/\sqrt{\mathrm{Hz}}$, so single-event integration over nanoseconds to microseconds achieves the charge sensitivities quoted above~\cite{GonzalezZalba2015}.

\paragraph{Thermal background and selection strategy}
At $T\!=\!\SI{4}{K}$, thermally populated modes peak near $f_{\mathrm{th}}\!\simeq\!k_B T/h\!\approx\!\SI{83}{GHz}$, but the Planck tail contributes significantly in the \SI{10}{-}\SI{30}{GHz} band. Thus, while resonance enhances transduction, thermal contamination demands aggressive selection. Our analysis (Sec.~\ref{sec:timing_psd}) uses (i) nanosecond-scale timing gates matched to the athermal burst and (ii) GHz-band PSD windows aligned with the PnC passband while suppressing thermal sidebands. The PnC also increases dwell time (raising signal) yet introduces stop bands to reject out-of-band thermal flux (lowering background).

\subsection{Illustrative scales and design levers}
Using Ge parameters at \SI{4}{K} (electrons: $D\!\approx\!\SI{13.5}{eV}$; holes: $D\!\approx\!\SI{4.5}{eV}$~\cite{YuCardona2010,Mahan2000}), PnC-enhanced $V_{\mathrm{mode}}$ (above), and $d\!=\!\SI{1}{\micro m}$, the model yields:
\begin{itemize}[leftmargin=*, itemsep=1pt]
  \item \textbf{Resonant band (\SI{10}{-}\SI{30}{GHz})}: $Q_{\mathrm{ind}}\!\sim\!10^{-3}$–$10^{-2}\,e$ per captured phonon.
  \item \textbf{Ballistic band ($\sim\!\SI{125}{GHz}$)}: $Q_{\mathrm{ind}}\!\gtrsim\!10^{-3}\,e$ owing to stronger $g(\omega)$ despite smaller $u_0$.
  \item \textbf{Levers}: reduce $d$; maximize $n_{\mathrm{eff}}$ and strain participation; place metal at displacement/strain nodes; tune $\omega_0$ (dot size/tension) into a low-background, high-transduction band; and optimize tank $Q$ and cryo-LNA (low-noise amplifier) noise for RF readout.
\end{itemize}

\subsection{Charge–phonon displacement trends}
\begin{figure}[htp]
  \centering
  \includegraphics[width=0.45\textwidth]{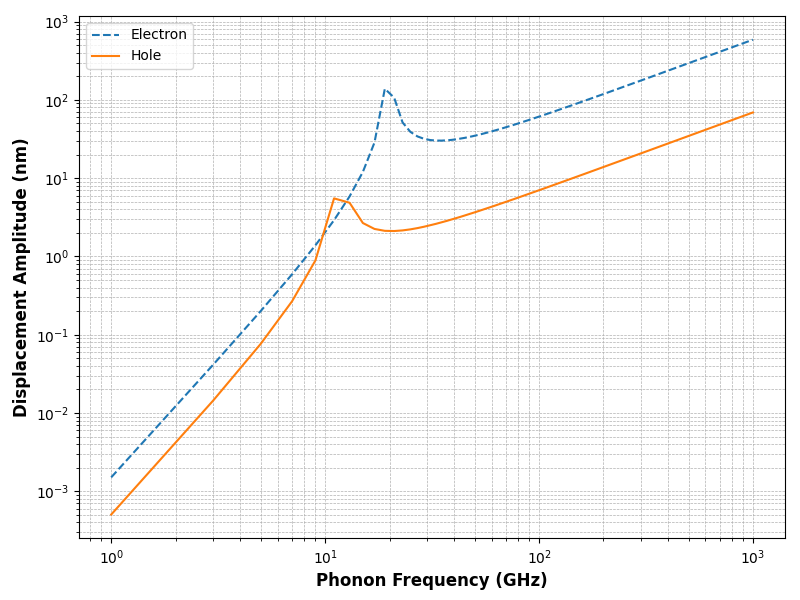}
  \caption{Calculated bound-carrier displacement $\delta x(\omega)$ for electrons and holes using the driven-oscillator model (Eq.~\ref{eq:dx}). A strong resonance appears near \SI{10}{-}\SI{30}{GHz}, with a high-frequency tail that remains appreciable into the ballistic band.}
  \label{fig:charge_displacement}
\end{figure}

Figure~\ref{fig:charge_displacement} shows the modeled carrier displacement $\delta x(\omega)$ using Eq.~\eqref{eq:dx}. A resonant enhancement appears around \SI{10}{-}\SI{30}{GHz} (device dependent). Above resonance, $\delta x$ decreases slowly with frequency because the growing $g(\omega)$ partially offsets the $u_0\!\propto\!\omega^{-1/2}$ scaling. Electrons exhibit larger displacements than holes across the band due to their larger $D$, suggesting an advantage for electron-like dots in Ge for ultimate sensitivity.

\paragraph{Caveats and outlook}
The oscillator model assumes isotropic confinement and a single resonance; real dots may exhibit anisotropy, multi-level structure, and crystal-field effects. Likewise, PnC dispersion and loss are geometry dependent. These effects will be refined with full 3D simulations and calibration data, but the simplified framework captures the dominant scalings and informs the selection windows and sensitivity projections used in Secs.~\ref{sec:timing_psd} and~\ref{sec:projected_sensitivity}.

\section{Prototype geometry and backgrounds}
\label{sec:prototype_backgrounds}

We consider a \SI{100}{g}-scale, high-purity Ge crystal (e.g., a right circular cylinder of \SIrange{2}{3}{cm} diameter and \SIrange{1}{1.5}{cm} thickness) with a near-surface PnC interaction layer that both \emph{slows} and \emph{guides} targeted acoustic modes and \emph{suppresses} thermally populated bands (see Figure~\ref{fig:gequlep_readout}). The functional PnC region comprises (i) a patterned surface layer with sub-wavelength features (depth \SIrange{0.1}{1}{\micro m}) and (ii) a \SIrange{1}{10}{\micro m} near-surface interaction zone that houses the dipole-defined quantum dots (QDs) and the QPC electrodes. The QPCs are positioned within \SIrange{50}{200}{\nano m} laterally and $\sim\!\SI{1}{\micro m}$ vertically from the active QD region to optimize capacitive coupling while limiting parasitic participation.

\paragraph{Dominant backgrounds and high-level mitigations}
At \SI{4}{K}, the principal backgrounds are:
\begin{itemize}
  \item \textbf{Thermal phonons}, characterized by $f_{\rm th}=k_BT/h\!\approx\!\SI{83}{GHz}$, with a nonzero Bose–Einstein tail in the \SI{10}{-}\SI{30}{GHz} signal band. \emph{Mitigations:} engineered PnC stop bands, GHz-band PSD selections, and sub-ns timing gates (Sec.~\ref{sec:timing_psd}).
  \item \textbf{Microphonics} coupling mechanical vibrations into the RF chain, producing low-frequency baseline motion and upconversion sidebands. \emph{Mitigations:} cryogenic mechanical isolation, rigid RF fixtures, notch/low-pass filtering of bias lines, and microphonic surveys of the dilution/fridge stack.
  \item \textbf{RF/electronics noise}, including cryo-LNA noise, QPC $1/f$ charge noise, and amplifier intermodulation. \emph{Mitigations:} narrow-band reflectometry (high-$Q$ tank), cross-spectrum readout, cold attenuators/isolators, and careful grounding/shielding.
\end{itemize}
Polished (\,$\eta_{\rm rms}\!\lesssim\!\SI{1}{nm}$\,) and passivated Ge surfaces preserve specular boundary conditions (Sec.~\ref{sec:reflectivity}), increasing geometric return and reducing conversion to lossy surface modes.

\subsection*{Stability and reproducibility of dipole-induced quantum dots}
Below \SI{10}{K}, shallow donors and acceptors freeze out to form dipole-defined QDs with reproducible spectra if impurity species and densities are well controlled. Crystal growth with dopant control at $\sim$\SI{1e10}{\per\cubic\centi\metre} (bulk) and patterned surface regions at $\sim$\SI{1e14}{\per\cubic\centi\metre} enable predictable localization depths. Post-growth thermal annealing and hydrogen passivation reduce charge noise and stabilize occupancy. Low-$T$ electron (or hole) tunneling spectroscopy and RF reflectometry are used to validate: (i) level spacing, (ii) charging energy, (iii) telegraph/random telegraph noise (RTN) rates, and (iv) device-to-device reproducibility.

\subsection*{PnC design, fabrication, and band engineering}
A surface PnC with lattice constant $a$ establishes Bragg stop bands near
\begin{equation}
  f_{\mathrm{gap}} \;\approx\; \frac{v_s}{2a},
  \label{eq:eq31}
\end{equation}
so with $v_s\!\approx\!\SI{5400}{m\,s^{-1}}$ and $a\!\approx\!\SI{25}{nm}$ one obtains $f_{\mathrm{gap}}\!\sim\!\SI{108}{GHz}$. This placement attenuates modes near and above $f_{\mathrm{th}}$ while leaving a pass band around \SI{10}{-}\SI{30}{GHz} for signal-rich athermal bursts; defect cavities within the PnC are tuned to \emph{slow} the pass-band modes (increase $n_{\rm eff}$), raising strain participation at the QD (Secs.~\ref{sec:collection}, \ref{sec:readout}). 

\emph{Process flow.} Sub-\SI{30}{nm} features are patterned in a hard mask (SiN or Cr) by electron-beam lithography (EBL) or deep-ultraviolet (DUV) photolithography, followed by anisotropic reactive ion etching (RIE) to transfer into Ge~\cite{Zhu2017,Bahari2019}. Focused-ion-beam (FIB) touch-up may trim unit cells to correct systematic offsets and sharpen the band edge. The patterned layer is only tens to hundreds of nm thick, preserving planarity and compatibility with QPC metallization. Throughout, CMP (chemical–mechanical polishing) and \emph{in situ} passivation maintain nm-scale roughness to keep the surface specularity parameter $p\!\to\!1$ (Sec.~\ref{sec:reflectivity}).

\subsection*{QPC sensitivity and cryogenic RF chain}
Single-phonon transduction demands sub-$10^{-3}\,e/\sqrt{\mathrm{Hz}}$ charge sensitivity. We embed a QPC in a cryogenic $LC$ tank and measure changes in the complex reflection coefficient at $\omega_{\rm RF}$~\cite{Field1993,Schoelkopf1998,Colless2013}. Strategies that have proven effective in GaAs and Si/SiGe translate to Ge:
\begin{itemize}[leftmargin=*,itemsep=2pt]
  \item Atomic layer deposition (ALD) AlO$_x$ (or HfO$_x$) passivation to suppress gate leakage and interface traps;
  \item High-$Q$ tank circuits with superconducting inductors and low-loss dielectrics to narrow the readout bandwidth;
  \item Cryogenic LNAs (4--8\,K) with adequate isolation; cold attenuators and isolators on the input to thermalize and prevent backaction;
  \item Cross-correlation of two RF chains to reject uncorrelated amplifier noise~\cite{GonzalezZalba2015,Chorley2012,Cassidy2007}.
\end{itemize}
Bias routing incorporates RC/LC filtering at multiple temperature stages to block microphonic upconversion and RF pickup.

\subsection*{Phonon lifetime, collection, and surface quality}
For high-purity Ge at sub-Kelvin temperatures, LA phonon lifetimes approach $\sim\!\SI{1}{\micro s}$ and cm-scale mean free paths have been reported in rare-event detectors~\cite{Agnese2014,Armengaud2017}. At \SI{4}{K}, intrinsic lifetimes are shorter but still compatible with multiple specular bounces and efficient geometric return (Sec.~\ref{sec:reflectivity}). The survival factor is
\[
\eta_{\rm prop}(\omega)=\exp\!\big[-n_d\,\sigma_{\rm abs}(\omega)\,L_{\rm eff}\big],
\]
with $n_d$ the (parasitic) dipole/defect density, $\sigma_{\rm abs}$ the DP-mediated absorption cross section, and $L_{\rm eff}$ the (multi-bounce) path length (Sec.~\ref{sec:collection}). Surface roughness and native oxides increase diffuse scattering; CMP, hydrogen annealing, and \emph{in situ} passivation restore high specularity and suppress conversion to lossy surface modes.

\subsection*{Spatial alignment of QDs and QPCs}
Maximal induced charge ($Q_{\rm ind}\!\propto\!\delta x/d$) requires small dot–QPC separation $d$ and controlled geometry (Sec.~\ref{sec:readout}). Sputter/ion implantation with rapid thermal annealing can localize shallow impurities with sub-\SI{100}{nm} lateral accuracy; alignment marks allow overlay of QPC metallization to within \SIrange{50}{200}{nm}. These CMOS-friendly steps are scalable to multi-pixel arrays for phonon imaging and coincidence selection (Sec.~\ref{sec:timing_psd}).

\subsection*{Background budget and selection hooks}
We translate the qualitative backgrounds into quantitative handles:
\begin{itemize}[leftmargin=*,itemsep=2pt]
  \item \textbf{Thermal phonons:} suppressed by a PnC stop band near \SI{100}{-}\SI{120}{GHz}, by PSD weighting that emphasizes \SI{10}{-}\SI{30}{GHz}, and by \SI{1}{ns} timing gates and multiplicity/coincidence triggers (Sec.~\ref{sec:timing_psd}).
  \item \textbf{Microphonics:} validated via \emph{shaker} and \emph{hammer} tests; line resonances notched; mechanical eigenmodes shifted away from $\omega_{\rm RF}$; cable strain-relief and potting at cold stages.
  \item \textbf{RF noise:} system noise temperature $T_{\rm sys}$ measured with hot/cold loads; cross-spectrum subtraction used during calibration runs; tank $Q$ optimized for the anticipated $Q_{\rm ind}$ spectrum.
\end{itemize}

\subsection*{Prototyping pathway}
As a first milestone, we envision a coupon-scale device (\SI{2}{cm} diameter, \SI{1}{cm} thickness; $\sim\!\SI{17}{g}$) with a single PnC port and one (or two) QPC(s):
\begin{itemize}
  \item Fabricate PnC (EBL/DUV + RIE + CMP/passivation); pattern QPCs and bias gates; deposit superconducting inductors for the RF tank.
  \item Optional SAW transducers at the perimeter to inject calibrated phonons and map PnC transfer.
  \item Commission at \SI{4}{K}: measure QPC noise floor, $Q_{\rm ind}$ vs.\ frequency using piezo-driven phonon injection, and verify timing/PSD selections with heater pulses.
  \item Iterate unit-cell geometry (FIB trim if needed) to sharpen the pass band and band edge; finalize selection thresholds and multiplicity settings.
\end{itemize}
These steps leverage established nanofabrication and RF-reflectometry techniques and are directly extensible to the \SI{100}{g} module.

\paragraph{Outlook}
While fabrication at \SI{25}{nm} pitch and cryogenic RF integration are nontrivial, all elements are within current capabilities. The combination of (i) high-specularity Ge surfaces, (ii) PnC slow-phonon engineering near \SI{10}{-}\SI{30}{GHz}, (iii) reproducible dipole-defined QDs, and (iv) RF-QPC readout with sub-$10^{-3}\,e$ sensitivity forms a credible path to single-phonon transduction with robust background rejection, enabling the physics reach quantified in Sec.~\ref{sec:projected_sensitivity}.


\section{Timing and PSD windows}
\label{sec:timing_psd}

Athermal phonons from rare, localized deposits arrive as a \emph{clustered burst}
following the anharmonic cascade (Sec.~\ref{sec:phonon_physics}), whereas thermal
phonons at \SI{4}{K} are \emph{stochastic} in time and broad in spectrum.
We therefore apply two complementary selections:
(i) a short \emph{timing gate} to capture correlated arrivals and
(ii) a frequency-domain \emph{PSD window} tuned to the signal-enhanced band while
suppressing thermal modes.

\subsection{Thermal vs.\ non-equilibrium phonons at \texorpdfstring{$4\,\mathrm{K}$}{4 K}}
The thermal occupation of a mode of frequency $f$ follows Bose--Einstein statistics,
\begin{equation}
  n(f,T) \;=\; \frac{1}{\exp(hf/k_BT)-1}.
\end{equation}
At $T=\SI{4}{K}$ the characteristic thermal frequency is
$f_{\mathrm{th}} \!\equiv\! k_B T/h \!\approx\! \SI{83}{GHz}$.
Thermal population in our signal band is \emph{reduced but nonzero}:
$n(\SI{10}{GHz},\SI{4}{K}) \!\approx\! 7.9$ and
$n(\SI{30}{GHz},\SI{4}{K}) \!\approx\! 2.3$,
so frequency alone cannot guarantee separation.
By contrast, a localized recoil produces a \emph{primary} THz phonon that
down-converts through several three-phonon generations; after $\sim\!4$--$6$
binary splits, the daughters populate the \SI{10}{-}\SI{30}{GHz} band with a
multiplicity $M\!\sim\!16$--$64$ arriving within a sub-ns envelope.
This temporal clustering is the key discriminant against the uncorrelated thermal flux.

\subsection{Timing gate design and multiplicity trigger}
The relevant timescale is set by the cascade duration and the flight time across
the sensing region. With group velocities $v_g\!\sim\!\SI{3.5}{-}\SI{5.4}{km\,s^{-1}}$,
a \SI{1}{ns} gate corresponds to a path of $\sim\!\SI{3.5}{-}\SI{5.4}{\micro m}$,
i.e., the spatial scale of the near-surface PnC/QW region. We adopt
\begin{equation}
  \Delta t_{\mathrm{gate}} \;\sim\; \SI{1}{ns} \quad (\text{typ. } 0.5{-}2~\mathrm{ns}),
\end{equation}
and require a \emph{cluster multiplicity} $N \ge 10$ above threshold within the gate.
A practical trigger is ``$N\!\ge\!N^\star$ in \SI{1}{ns}'' with $N^\star$ chosen to yield
a low false-alarm probability under thermal rates (see below). In multi-sensor layouts,
a \mbox{2-of-4} spatial coincidence within $\Delta t_{\mathrm{coinc}}\!\sim\!\SI{2}{ns}$
further suppresses randoms and leverages anisotropic phonon focusing in Ge for localization.

\subsection{PSD window and matched filtering}
We define a spectral weighting $W(f)$ that emphasizes the signal-rich band while
discounting thermal sidebands:
\begin{equation}
  W(f) \;=\;
  \begin{cases}
    w_1, & f\in[\SI{10}{GHz},\,\SI{30}{GHz}] \\
    w_2\ll w_1, & \text{elsewhere \ (with optional notches)}
  \end{cases}
\end{equation}
and compute the windowed PSD or bandpower over the gate,
$P_{\mathrm{band}}=\int |X(f)|^2 W(f)\,df$, where $X(f)$ is the
demodulated QPC signal. A simple implementation uses a bank of I/Q bandpass
filters (digital or analog) centered across \SI{10}{-}\SI{30}{GHz}, combined
with a short matched filter to the expected burst envelope. This yields a
scalar test statistic per gate to feed into the multiplicity/coincidence logic.

\subsection{Expected thermal count and false-alarm control}
Let $\kappa_{\mathrm{th}}(f)$ denote the effective thermal \emph{arrival rate
density} in the readout band at the sensor after geometric acceptance, PnC
transfer, and detector response. The expected thermal count within a gate is
\begin{equation}
  \mu_{\mathrm{th}} \;=\; \Delta t_{\mathrm{gate}} \int {\mathcal{B} \kappa_{\mathrm{th}}(f)\,df},
  \qquad \mathcal{B}=[\SI{10}{GHz},\,\SI{30}{GHz}].
\end{equation}
For Poisson statistics, the probability of observing $N$ $\ge N^\star$ thermal
hits in a single gate is
$P_{\mathrm{FA}} = 1 - \sum_{k=0}^{N^\star-1} e^{-\mu_{\mathrm{th}}}\mu_{\mathrm{th}}^{k}/k!$.
Choosing $N^\star$ such that $P_{\mathrm{FA}}\!\ll\!R_{\mathrm{gate}}^{-1}$ (with
$R_{\mathrm{gate}}$ the gate rate) controls the global false-alarm budget.
Because signal bursts deliver $M\!\sim\!16$--$64$ daughters \emph{within}
\SI{1}{ns}, the same selection is highly efficient for rare-event phonons.

\subsection{PnC-assisted spectral filtering}
The near-surface PnC provides hardware-level spectral shaping. A periodic
array with lattice constant $a$ produces a Bragg stop band near
  $f_{\mathrm{gap}} \;\approx\; \frac{v_s}{2a}$ (see Eq.\ref{eq:eq31}.),
so with $v_s\!\approx\!\SI{5400}{m\,s^{-1}}$ and $a\!\approx\!\SI{25}{nm}$ one
obtains $f_{\mathrm{gap}}\!\sim\!\SI{108}{GHz}$, suppressing thermally
populated modes near and above $f_{\mathrm{th}}$ while leaving the
\SI{10}{-}\SI{30}{GHz} passband intact~\cite{Maldovan2013,Joannopoulos2008}.
Defect modes within the PnC cavity can be tuned to slow and confine the
target band, increasing dwell time and strain participation at the dot (and
thereby transduction efficiency), while out-of-band thermal flux is reflected
or scattered before reaching the sensor.

\subsection{Directionality and multi-channel imaging}
Ballistic phonons focus along high-symmetry axes in Ge. A sparse array of
QPC pixels or multiple PnC ports enables coarse phonon imaging:
relative arrival times and bandpowers across channels reconstruct the
incident direction. Requiring geometric consistency between a hypothesized
source point and measured channel times within $\Delta t_{\mathrm{coinc}}$
adds a powerful topological veto against random thermal hits.

\subsection{Illustrative PSD contrast at \texorpdfstring{$4\,\mathrm{K}$}{4 K}}
\begin{figure}[htp]
  \centering
  \includegraphics[width=0.85\linewidth]{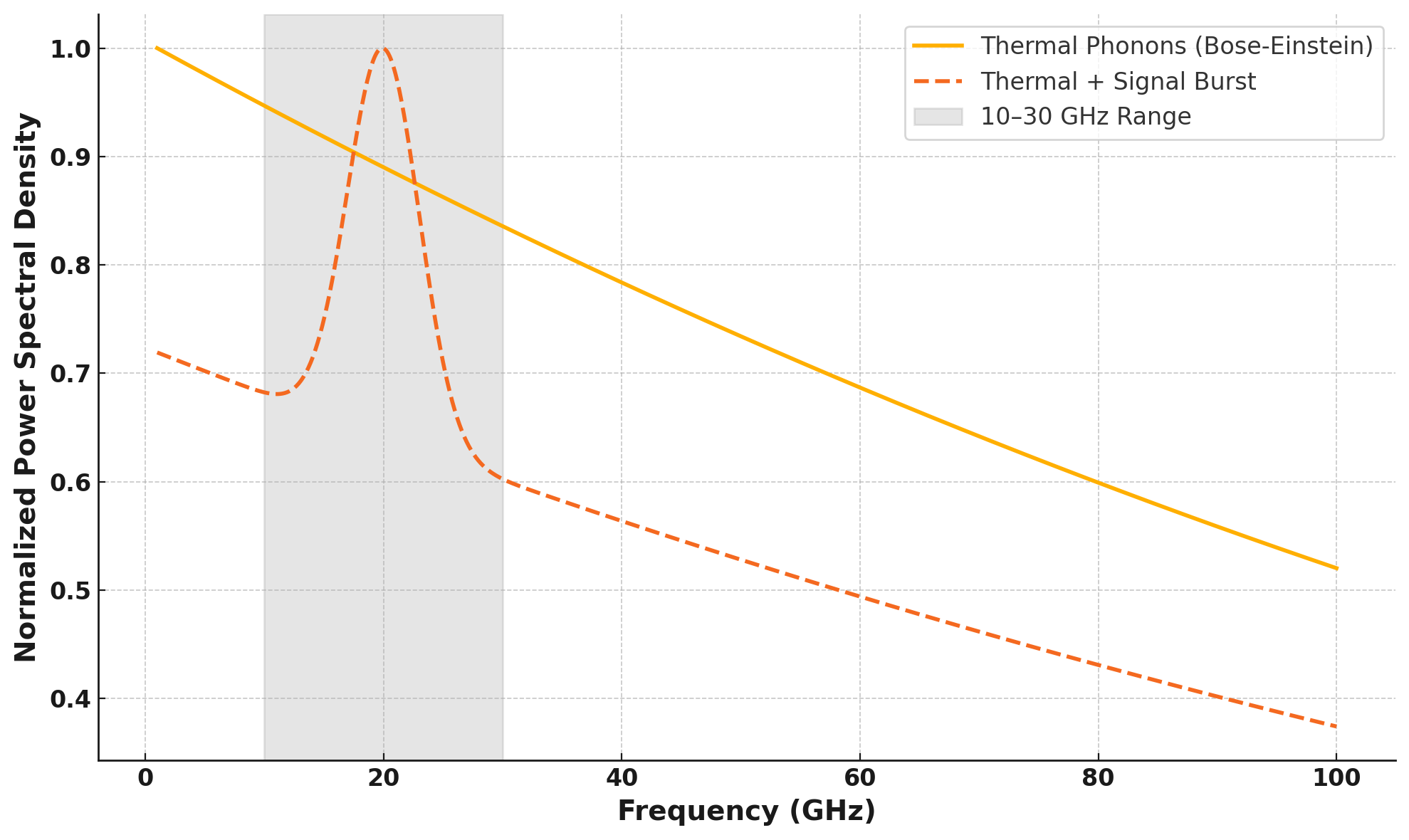}
  \caption{Normalized phonon PSD in Ge at \SI{4}{K}. Solid: thermal background $\propto n(f,T)$.
  Dashed: thermal plus a burst centered at \SI{20}{GHz} (width \SI{3}{GHz}) from an anharmonic cascade.
  The shaded \SI{10}{-}\SI{30}{GHz} band is used for selection (with matched filters and multiplicity/coincidence).}
  \label{fig:phonon_psd}
\end{figure}

Figure~\ref{fig:phonon_psd} compares the normalized thermal PSD,
proportional to $n(f,T)$, with a simulated Gaussian burst centered at
\SI{20}{GHz} (width \SI{3}{GHz}) representative of a down-converted daughter
population. While the burst sits within the thermal continuum, the
\SI{10}{-}\SI{30}{GHz} window (shaded) exhibits a localized excess when
integrated over a \SI{1}{ns} gate and combined with multiplicity and
coincidence requirements.

\paragraph{Operational summary}
We implement a \SI{1}{ns} gate with a band-limited, matched-filter PSD statistic
in \SI{10}{-}\SI{30}{GHz}, require $N\!\ge\!N^\star$ hits (set by a global
$P_{\mathrm{FA}}$ target), and apply a \mbox{2-of-4} channel coincidence with
directional consistency. The PnC cavity provides a hardware stop band near
\SI{100}{-}\SI{120}{GHz} and slow-phonon enhancement in the passband.
Together these measures suppress thermal backgrounds while retaining
high efficiency for the compact athermal phonon bursts produced by rare,
low-energy events.


\section{Projected sensitivity}
\label{sec:projected_sensitivity}

We estimate the discovery-limited sensitivity of a single \SI{100}{g} Ge module operated at \SI{4}{K}. Signal acceptances are taken from the efficiency pipeline in Secs.~\ref{sec:collection}--\ref{sec:readout}, together with the timing and PSD selections of Sec.~\ref{sec:timing_psd}. Unless stated otherwise, we assume the \emph{standard halo model} with local density $\rho_\chi=\SI{0.3}{GeV\,cm^{-3}}$, a Maxwellian speed distribution with $v_0=\SI{220}{km\,s^{-1}}$, Galactic escape speed $v_{\rm esc}=\SI{544}{km\,s^{-1}}$, and Earth speed $v_E=\SI{232}{km\,s^{-1}}$~\cite{Essig2012,Battaglieri2017}. We treat DM--electron scattering in the two benchmark mediator scenarios of Refs.~\cite{Essig2012,Essig2016}: a heavy mediator with $F_{\rm DM}(q)=1$ and a light mediator with $F_{\rm DM}(q)=(\alpha m_e/q)^2$. The semiconductor material response follows the formalism of~\cite{Essig2012,Bloch2017,Knapen2021}.

\paragraph{Assumptions}
The phonon\,\textrightarrow\,charge transducer yields a \emph{selection-corrected} threshold in the $\SI{1e-3}{eV}$--$\SI{1e-2}{eV}$ range (Sec.~\ref{sec:readout}), with overall athermal-phonon collection efficiency $\eta_{\rm collection}\!\sim\!0.9$ and multiplicity-assisted primary-phonon detection $\eta_{\rm primary}\!\sim\!0.8$ (Sec.~\ref{sec:collection}). A \SI{1}{ns} timing gate and \SI{10}{-}\SI{30}{GHz} PSD window (Sec.~\ref{sec:timing_psd}) keep the search background-free up to $\mathcal{O}(0.1)$~kg$\cdot$yr exposure per module; at larger exposures, the solar CE$\nu$NS background sets the ultimate floor~\cite{Billard2014}.

\paragraph{Rate and limit construction}
For DM--$e^-$ scattering, the expected rate per unit target mass can be written schematically as
\begin{equation}
  R(m_\chi,\bar{\sigma}_e) \;=\; \frac{\rho_\chi}{m_\chi}\,\bar{\sigma}_e
  \int\!dq\,d\omega\; |F_{\rm DM}(q)|^2\,
  \mathcal{M}(q,\omega)\,\eta(v_{\min})\,\varepsilon(\omega)\,,
\end{equation}
where $\bar{\sigma}_e$ is the reference free-electron cross section (defined at $q_0=\alpha m_e$), $\mathcal{M}(q,\omega)$ encodes the crystal/atomic response (including the phonon-mediated channel and ionization/phonon yields), $\eta(v_{\min})$ is the mean inverse speed (halo integral), and $\varepsilon(\omega)$ is the total selection efficiency~\cite{Essig2012,Essig2016,Bloch2017}. When comparing to nuclear–recoil limits quoted in terms of the spin–independent per–nucleon cross section $\sigma_p$, the electron and nucleon benchmarks are related by simple reduced-mass scalings that depend on the mediator regime. 

\emph{Light mediator ($F_{\rm DM}\propto q^{-2}$).} Under the usual assumptions (equal mediator couplings to electrons and nucleons, $f_e=f_p=f_n$, and coherent nuclear scattering), one finds~\cite{Mei2018}
\begin{equation}
\label{eq:map_light}
\bar{\sigma}_e \;=\; 4\,\frac{\mu_{\chi e}^{2}}{\mu_{\chi N}^{2}}\,\sigma_{p}\,,
\end{equation}
with $\mu_{\chi e}=m_\chi m_e/(m_\chi+m_e)$ the $\chi$--$e$ reduced mass and $\mu_{\chi N}=m_\chi m_N/(m_\chi+m_N)$ the $\chi$--nucleus reduced mass (for Ge, $m_N\simeq A\, m_p$). 

\emph{Heavy mediator / contact limit ($F_{\rm DM}=1$).} If the same heavy mediator couples with equal strength to electrons and protons ($f_e=f_p=f_n$), the reference cross sections are related by~\cite{Mei2018}
\begin{equation}
\label{eq:map_heavy}
\bar{\sigma}_e \;=\; \left(\frac{\mu_{\chi e}}{\mu_{\chi p}}\right)^{2}\,\sigma_{p}\,,
\end{equation}
with $\mu_{\chi p}=m_\chi m_p/(m_\chi+m_p)$ the $\chi$--proton reduced mass. The coherent enhancement to the \emph{nuclear} cross section is then accounted for separately via 
\begin{equation}
  \sigma_0 \;=\; \sigma_p\,\frac{\mu_{\chi N}^2}{\mu_{\chi p}^2}\,A^2\,,
\end{equation}
which enters the nuclear recoil rate. Detailed derivations of Eqs.~\eqref{eq:map_light}–\eqref{eq:map_heavy} and the associated rate expressions can be found in Mei \emph{et~al.}~\cite{Mei2018}. 

In the background-free regime, a one-sided \SI{90}{\percent}~CL upper limit follows from $N_{90}\simeq 2.3 = R\,\mathcal{E}$ with exposure $\mathcal{E}$ (kg$\cdot$yr). For larger exposures we employ a profile-likelihood including residual thermal and CE$\nu$NS backgrounds~\cite{Billard2014}.

\begin{figure}[h]
  \centering
  \includegraphics[width=0.9\linewidth]{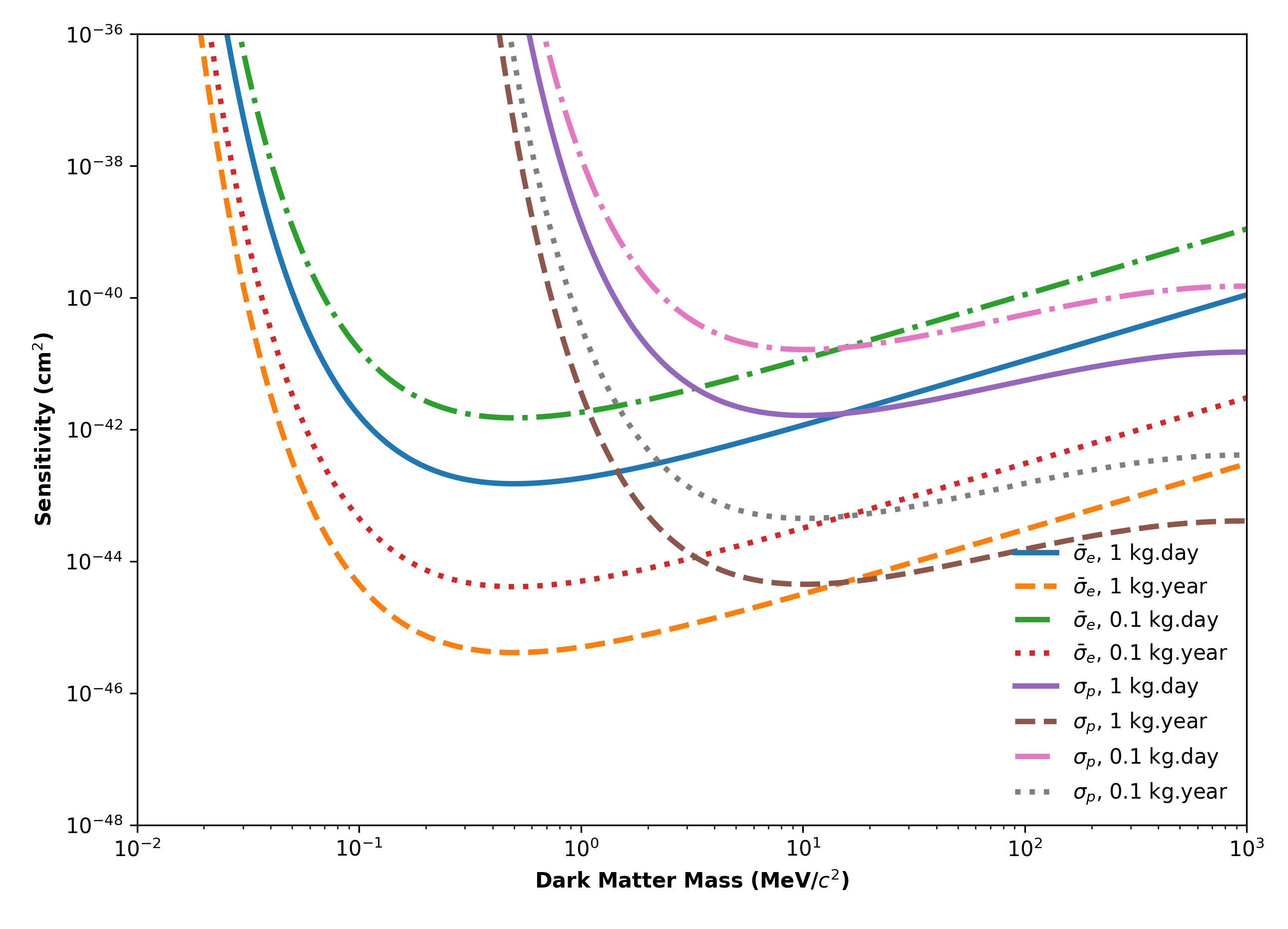}\\[4pt]
  \includegraphics[width=0.9\linewidth]{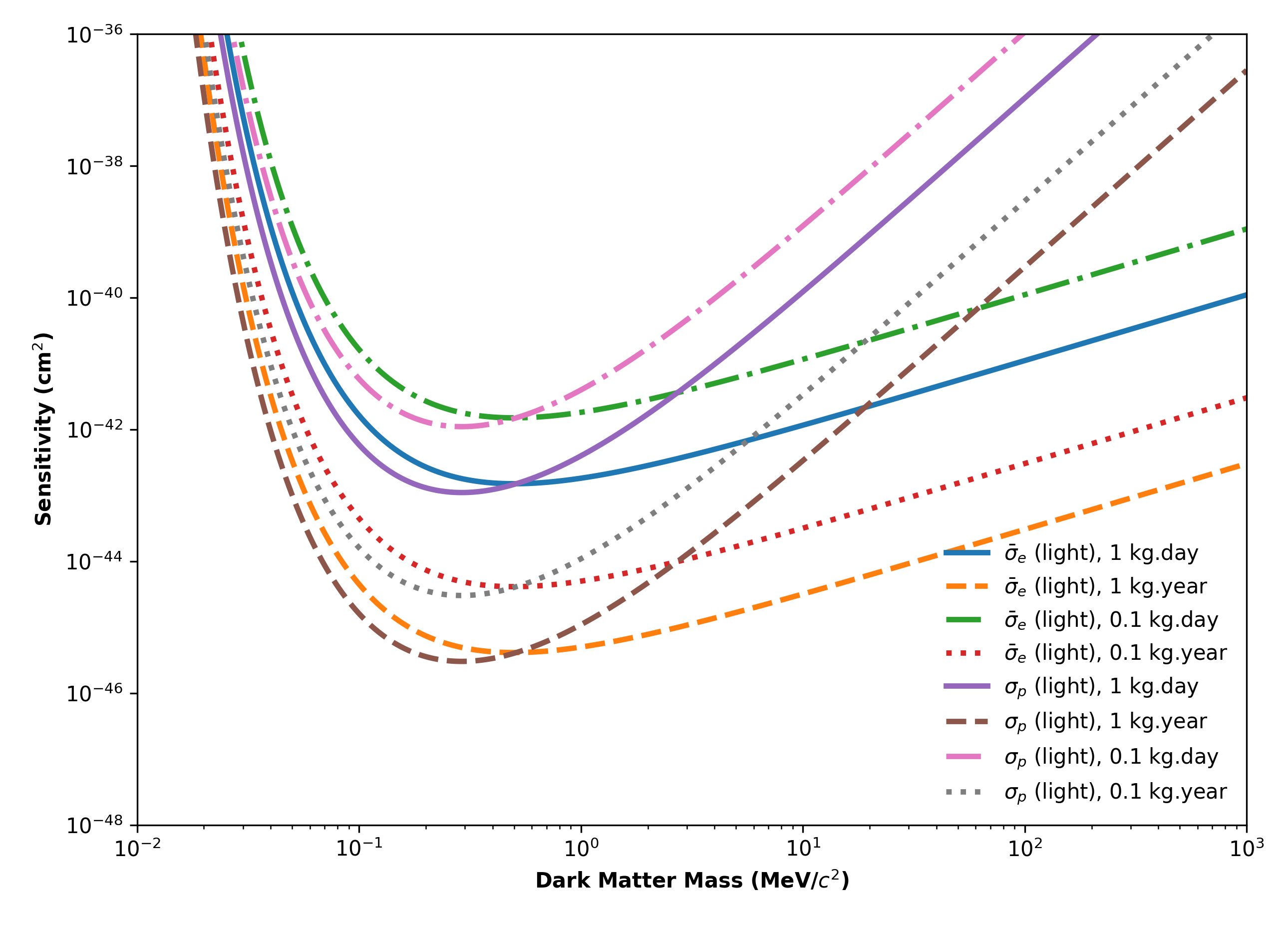}
  \caption{Projected 90\%~CL (Poisson, 2.3 counts; background-free) limits on the DM--electron reference cross section $\bar{\sigma}_e$ and the DM--nucleon cross section $\sigma_p$ for a Ge target. Top: heavy mediator ($F_{\rm DM}=1$). Bottom: light mediator ($F_{\rm DM}\propto 1/q^{2}$). Each panel shows projected sensitivities for exposures of \SI{0.1}{kg\cdot day}, \SI{1}{kg\cdot day}, \SI{0.1}{kg\cdot yr}, and \SI{1}{kg\cdot yr}, with selection efficiencies and thresholds folded in as described in the text.
 The overall normalization is intended for design-level comparison; for complete rate calculations and form-factor treatments see Refs.~\cite{Essig2012,Essig2016,Bloch2017,Knapen2021}. For the thresholds and sub-kg\,yr exposures considered in this figure, solar pp and $^7$Be CE$\nu$NS event rates lie below the discovery curves. As exposure approaches $\gtrsim \mathbf{1}\ \mathrm{kg\,yr}$ with the same threshold, CE$\nu$NS begins to shape the sensitivity floor; this regime is between 10$^{-49}$cm$^2$ to 10$^{-47}$cm$^2$ corresponding to a DM mass range of 1 - 100 MeV/c$^2$.}
  \label{fig:proj_sensitivity}
\end{figure}

\paragraph{Projected reach (sub-MeV)}
For a single \SI{100}{g} module operating for one year ($\mathcal{E}=\SI{0.1}{kg\cdot yr}$), the sensitivity extends into the sub-MeV mass range, with the turn-on governed by the phonon threshold and the \SI{10}{-}\SI{30}{GHz} passband. Scaling to \SI{1}{kg\cdot yr} improves reach by roughly an order of magnitude in cross section. Figure~\ref{fig:proj_sensitivity} shows the projections for the heavy- and light-mediator benchmarks. The shape reflects the threshold-limited rise below $\sim\!\SI{0.3}{MeV}/c^2$ and the mediator form factor at low $q$~\cite{Essig2012,Essig2016,Knapen2021}. Final numbers depend on $\eta_{\rm collection}$, multiplicity $M$, the athermal efficiency $\eta_{\rm ath}$, and the residual background rate; these are propagated as nuisance parameters in the limit setting. 

For the heavy-mediator benchmark ($F_{\rm DM}=1$), the differential rate is effectively $q$-independent (contact interaction). Once the selection efficiency saturates in our analysis, the dominant explicit mass dependence of the predicted yield arises from the flux factor $\rho_\chi/m_\chi$ in the rate integral. As a result, the inferred limit on the proton cross section scales only weakly with mass, $\sigma_p^{\rm lim}\propto m_\chi$, which appears nearly horizontal (“flattened”) on log–log axes for $m_\chi \gtrsim 100~\mathrm{MeV}$. In contrast, for the light-mediator scenario the integrand retains strong $q$-dependence through $|F_{\rm DM}(q)|^2 \propto q^{-4}$, preserving a pronounced $m_\chi$ dependence and preventing such flattening. 

\paragraph{Systematics and neutrino floor}
The dominant systematics in the sub-MeV regime arise from (i) the effective energy threshold and selection efficiency $\varepsilon(\omega)$; (ii) the slow-phonon quality factor and mode volume of the PnC cavity (entering transduction and acceptance); and (iii) bulk defect densities affecting phonon propagation (Sec.~\ref{sec:collection}). For exposures $\gtrsim\!\mathrm{kg}\cdot\mathrm{yr}$, solar $pp$ and $^7$Be CE$\nu$NS backgrounds enter the sensitivity budget, setting the discovery floor in the \SI{1}{-}\SI{100}{MeV}/$c^2$ mass range~\cite{Billard2014}.

\section{Systematics and calibration}
\label{sec:systematics_calibration}
We quantify and calibrate the leading detector–response uncertainties that impact the projected reach of Sec.~\ref{sec:projected_sensitivity}. The dominant effects are:
(i) \emph{selection threshold and baseline noise}, (ii) \emph{PnC band-edge placement and mode $Q$}, (iii) \emph{bulk defect density $n_d$ and the associated absorption cross section $\sigma_{\rm abs}(\omega)$}, (iv) \emph{surface/boundary losses}, and (v) \emph{RF-QPC gain, linearity, and stability}. For each category, the corresponding calibration handles are summarized below, drawing on established methods in cryogenic phonon detection, semiconductor acoustics, and RF reflectometry~\cite{SwartzPohl1989,Klitsner1988,Joannopoulos2008,Maldovan2013,Field1993,Schoelkopf1998,Colless2013,GonzalezZalba2015,Agnese2014,Armengaud2017,Mahan2000,YuCardona2010}.
\subsection{Threshold and baseline noise}
The selection efficiency as a function of deposited (phonon) energy is modeled by
\begin{equation}
  \varepsilon(E)=\frac{1}{2}\,\mathrm{erfc}\!\left(\frac{E_{\rm thr}-E}{\sqrt{2}\,\sigma_E}\right),
  \label{eq:turnon}
\end{equation}
with effective threshold $E_{\rm thr}$ and baseline energy resolution $\sigma_E$. These parameters are constrained by three complementary measurements:
\begin{itemize}[leftmargin=*,itemsep=2pt]
  \item \textbf{Random/blank triggers} recorded without injected signals to determine the baseline power spectral density and $\sigma_E$ under the timing/PSD windows of Sec.~\ref{sec:timing_psd}.
  \item \textbf{Heater pulses} delivered by a thin-film resistor with calibrated current drive; the injected energy $\Delta E=\!\int I^2(t)R\,dt$ provides absolute anchors for $\varepsilon(E)$ near threshold, as standard in cryogenic phonon calorimetry~\cite{Agnese2014,Armengaud2017}.
  \item \textbf{Narrowband phonon injection} using an on-chip piezo/Interdigital Transducer (IDT) or Surface Acoustic Wave (SAW) emitter to deliver wave packets in the \SI{10}{-}\SI{30}{GHz} band and to map the matched-filter response of the RF-QPC chain (Sec.~\ref{sec:timing_psd}).
\end{itemize}
In the likelihood for $\sigma_e$, $(E_{\rm thr},\sigma_E)$ enter as Gaussian-constrained nuisances; fractional shifts $\delta E_{\rm thr}/E_{\rm thr}\!\sim\!5$--$10\%$ typically control the low-mass turn-on for the heavy-mediator benchmark~\cite{Essig2012}.

\subsection{PnC band-edge placement and mode $Q$}
The placement of the slow-phonon pass band and the cavity quality factor set the strain participation and hence the phonon-to-charge transduction (Sec.~\ref{sec:readout}). Two measurements are used:
\begin{itemize}[leftmargin=*,itemsep=2pt]
  \item \textbf{Dispersion and band edges} from S-parameter transmission between an injector and the QPC port while sweeping frequency; comparison with Finite Element Method (FEM) and Plane-Wave Expansion (PWE) simulations yields the band diagram and group velocity $v_g(f)$~\cite{Joannopoulos2008,Maldovan2013}.
  \item \textbf{Ringdown $Q$} from pulsed excitation of a defect mode with exponential decay $Q=\pi f\,\tau_{\rm ring}$; temperature scans separate intrinsic from boundary-limited loss channels.
\end{itemize}
Priors on the pass-band centroid ($\pm\SI{1}{GHz}$) and on $Q$ (20--30\%) are updated by these measurements and propagate into the acceptance $\varepsilon(\omega)$ through the slow-phonon enhancement and the PSD window (Sec.~\ref{sec:timing_psd}).

\subsection{Bulk defects and propagation survival}
Propagation losses are described by Beer–Lambert attenuation,
$\eta_{\rm prop}(\omega)=\exp[-n_d\,\sigma_{\rm abs}(\omega)\,L_{\rm eff}]$ (Sec.~\ref{sec:collection}). The ingredients are constrained as follows:
\begin{itemize}[leftmargin=*,itemsep=2pt]
  \item \textbf{Effective path \boldmath$L_{\rm eff}$ and attenuation length} from time-of-flight (TOF) measurements with impulsive sources; multi-bounce tails constrain $L_{\rm eff}$ and the attenuation coefficient $\alpha(\omega)$~\cite{SwartzPohl1989,Klitsner1988}.
  \item \textbf{Defect density \boldmath$n_d$ and DP absorption} from independent materials characterization (EPR/PL, low-temperature transport) and from the frequency dependence of $\alpha(\omega)$ around the PnC pass band, informing $\sigma_{\rm abs}(\omega)$ via deformation-potential theory~\cite{Mahan2000,YuCardona2010}.
\end{itemize}
A log-normal prior on $\eta_{\rm prop}$ (typical median $0.9$ with 10--15\% width) is used, and projections in Fig.~\ref{fig:proj_sensitivity} are re-evaluated under draws from this prior.

\subsection{Surface/boundary losses}
Specular reflection at Ge–vacuum boundaries is expected ($R\!\to\!1$), whereas roughness and native oxides introduce diffuse scattering and mode conversion~\cite{SwartzPohl1989,Klitsner1988}. A specularity parameter $p$ (probability of specular reflection) is extracted by:
\begin{itemize}[leftmargin=*,itemsep=2pt]
  \item comparing TOF tail shapes and echo energies before/after CMP and \emph{in situ} passivation, and
  \item varying grazing angle via injector/receiver geometry to probe the angular dependence of boundary scattering.
\end{itemize}
The parameter $p$ modifies $\eta_{\rm geom}$ (Sec.~\ref{sec:collection}); polished, passivated surfaces routinely yield $p\!\to\!1$ at cryogenic temperatures~\cite{SwartzPohl1989}.

\subsection{RF-QPC gain, linearity, and absolute scale}
The RF reflectometry chain converts induced charge to voltage at the demodulator. The following calibrations are performed:
\begin{itemize}[leftmargin=*,itemsep=2pt]
  \item \textbf{Charge responsivity} $dV/dQ$ from single-electron steps in a nearby dot or controlled gate-charge excursions; Coulomb-diamond slopes provide absolute charge referencing~\cite{Field1993,Schoelkopf1998,GonzalezZalba2015,Colless2013}.
  \item \textbf{Noise temperature} $T_{\rm sys}$ via hot/cold loads and cross-spectrum readout (two amplifiers on the same device) to suppress uncorrelated LNA noise~\cite{Colless2013}.
  \item \textbf{Linearity and compression} from RF-power sweeps, monitoring $dV/dQ$ and intermodulation products; operation is maintained $>10$\,dB below the 1\,dB compression point.
\end{itemize}
These measurements fix the conversion from $\delta x$ (Sec.~\ref{sec:readout}) to signal-to-noise ratio (SNR) and, through Eq.~\eqref{eq:turnon}, to the turn-on $\varepsilon(E)$.

\subsection{Recoil-type tagging: neutron and photoexcitation calibrations}
Data-like validation uses tagged samples:
\begin{itemize}[leftmargin=*,itemsep=2pt]
  \item \textbf{Neutron tags} (using a Deuterium-Deuterium (DD) neutron generator or an AmBe with moderators as a neutron source) to produce nuclear recoils and benchmark timing/PSD response against CE$\nu$NS-like phonon bursts~\cite{Agnese2014,Armengaud2017}.
  \item \textbf{Photoexcitation tags} (near-IR LEDs/lasers or x-ray sources) to produce electronic recoils; distinct phononization pathways supply an ER control sample~\cite{Agnese2014}.
\end{itemize}
Tagged datasets are analyzed identically to WIMP-search data to obtain data-driven efficiencies and to validate the background model used for Fig.~\ref{fig:proj_sensitivity}.

\subsection{Blinding and stability monitoring}
Time-based blinding masks are applied to search windows. Run-by-run stability is tracked with interleaved heater/phonon injections, continuous baselines (random triggers), and periodic $dV/dQ$ checks. Slow drifts (temperature, bias) are corrected; residuals are incorporated via nuisance priors.

\begin{table}[t]
  \centering
  \footnotesize
  \setlength{\tabcolsep}{3pt}
  \renewcommand{\arraystretch}{1.15}
  \caption{Systematic parameters, calibration handles, and representative priors used in the reach projection. Prior widths are updated by the dedicated calibrations listed.}
  \begin{tabular}{@{}p{0.29\linewidth} p{0.40\linewidth} p{0.29\linewidth}@{}}
    \toprule
    \textbf{Parameter} & \textbf{Calibration handle} & \textbf{Prior (illustrative)} \\
    \midrule
    $E_{\rm thr},\,\sigma_E$ & blanks, heater pulses, narrowband injection & Gaussian, 5--10\% \\
    PnC band center, $Q$ & S-parameter sweep, ringdown ($Q=\pi f\tau$) & Gaussian, 10--30\% \\
    $\eta_{\rm prop}$ ($n_d,\sigma_{\rm abs}$) & TOF attenuation, EPR/PL, DP fits & Log-normal, 10--15\% \\
    Boundary specularity $p$ & TOF echoes pre/post CMP/passivation & Gaussian, 5--10\% \\
    $dV/dQ$, $T_{\rm sys}$ & charge steps, hot/cold loads, cross-spectrum & Gaussian, 5--10\% \\
    ER/NR selection & neutron \& photo tags (timing/PSD) & Binomial, data-driven \\
    \bottomrule
  \end{tabular}
\end{table}

With these calibrations, acceptance and noise uncertainties are constrained \emph{in situ}, stabilizing the selection efficiency that governs the low-mass turn-on and ensuring that the projected limits in Fig.~\ref{fig:proj_sensitivity} are robust to reasonable variations in device and materials parameters.


\section{Experimental roadmap and risk mitigation}
\label{sec:roadmap}

We structure the experimental program in three stages that progress from device-physics validation to array operation. Each stage specifies quantitative performance benchmarks and calibration measurements that feed into the systematics framework of Sec.~\ref{sec:systematics_calibration}.

\subsection*{Stage I—Coupon validation (cm scale, 4--10\,K)}
This stage establishes slow-phonon engineering, transduction, and thermal rejection on cm-scale Ge coupons. Phononic-crystal (PnC) unit cells with sub-\SI{30}{nm} pitch are fabricated on Ge (EBL/DUV and RIE with SiN/Cr masks) and characterized by transmission and ringdown to extract dispersion relations and defect-mode quality factors; measurements are compared with FEM/PWE designs~\cite{Joannopoulos2008,Maldovan2013,Zhu2017,Bahari2019}. Narrowband phonon injection via on-chip piezo/IDT or SAW emitters provides wave packets in the \SI{10}{-}\SI{30}{GHz} band for time-of-flight (TOF) and attenuation studies, constraining $\eta_{\rm prop}(\omega)$ and boundary specularity~\cite{SwartzPohl1989,Klitsner1988}. The charge transduction chain is calibrated by embedding a QPC in a cryogenic tank circuit and determining $dV/dQ$, noise temperature, and linearity from gate-charge steps and cross-spectrum readout~\cite{Field1993,Schoelkopf1998,Colless2013,GonzalezZalba2015}. Metal-film heaters and narrowband injection anchor the turn-on $\varepsilon(E)$ and validate matched filters (Sec.~\ref{sec:systematics_calibration}). Performance benchmarks for this stage are: pass-band center within $\pm\SI{1}{GHz}$ of design, defect-mode $Q\!\ge\!5\times 10^3$, baseline charge sensitivity $<\!10^{-3}\,e/\sqrt{\mathrm{Hz}}$, timing resolution $\le\!\SI{1}{ns}$, and a thermal-rejection factor $>\!10^3$ in the analysis band.

\subsection*{Stage II—Module demonstration (\SI{10}{-}\SI{100}{g}, 4\,K)}
This stage demonstrates selection-corrected thresholds, background controls, and stability in an integrated module. Devices combine near-surface PnCs with dipole-defined quantum dots and proximate QPCs (50--\SI{200}{nm} lateral separation); CMP and \emph{in situ} passivation preserve high specularity (Sec.~\ref{sec:reflectivity}). A background-control program addresses microphonics (survey, line notching, vibration isolation) and RF/electronic pickup; ER/NR tagging with LEDs/x-rays and neutrons validates timing/PSD selections~\cite{Agnese2014,Armengaud2017}. Blinding, interleaved injections, and run-by-run gain checks track $(E_{\rm thr},\sigma_E)$ and $dV/dQ$ (Sec.~\ref{sec:systematics_calibration}). Benchmarks for this stage are: a selection-corrected threshold in the $\SI{1e-3}{eV}$--$\SI{1e-2}{eV}$ range, background-free exposure $\ge\!0.05$\,kg$\cdot$yr, and data-driven efficiencies consistent with projections at the 10\% level.

\subsection*{Stage III—Scaling and array integration ($\sim$kg, 4\,K)}
The final stage scales to multi-module arrays with multiplexed RF readout and uniform acceptance, targeting $\mathcal{O}(1)$\,kg$\cdot$yr exposures and approaching the solar CE$\nu$NS floor~\cite{Billard2014}. The array architecture employs cold combining or frequency-division multiplexing of QPC tanks with cross-talk control and per-channel calibration. Wafer-level PnC/QPC process control (test coupons per wafer) and per-module dispersion/$Q$ verification enable automated updates of nuisance priors in the global likelihood. Low-background integration follows cryogenic Ge detector practice using radio-clean materials and a shielded cryostat~\cite{Agnese2014,Armengaud2017}. Target array benchmarks include stable multi-module operation, fractional threshold spread $<\!15\%$, live-time fraction $>\!80\%$, and combined sensitivity consistent with design within the systematic envelope.

\subsection*{Facilities and cadence}
Measurements draw on university nanofabrication (EBL/DUV, RIE, ALD), cryogenic test stands at 4\,K, RF metrology, and neutron/photonic calibration sources. The three stages proceed sequentially, with feedback from each stage informing device refinements and analysis priors.

\subsection*{Anticipated technical risks and controls}
Several failure modes are considered together with control measures. (i) \emph{PnC fabrication yield and band placement:} design-of-experiments for unit-cell geometries, hard masks (SiN/Cr), and post-fabrication focused-ion-beam trim; broader pass bands at higher drive serve as fallbacks~\cite{Zhu2017,Bahari2019,Joannopoulos2008}. (ii) \emph{Boundary losses from roughness/oxide:} CMP to nm roughness, \emph{in situ} passivation, and TOF-echo monitoring with re-polish as needed~\cite{SwartzPohl1989,Klitsner1988}. (iii) \emph{Bulk defects reducing $\eta_{\rm prop}$:} tighter feedstock specifications, EPR/PL screening, annealing, and lower-temperature operation to lengthen lifetimes~\cite{Agnese2014,Armengaud2017}. (iv) \emph{QPC noise and gain instability:} ALD AlO$_x$ interfaces, cross-spectrum readout, periodic $dV/dQ$ calibration, and operation $>\!10$\,dB below compression~\cite{Field1993,Schoelkopf1998,Colless2013,GonzalezZalba2015}. (v) \emph{Microphonics and RF pickup:} mechanical isolation, strain-relieved cabling, cold attenuators/isolators, and notch filters characterized with shaker/hammer runs. (vi) \emph{Execution risks (e.g., tool availability, staffing):} cross-institutional staffing, shared characterization with partner laboratories, parallel fabrication/measurement streams, and a rolling set of test coupons.

\paragraph{Integration with analysis}
Each stage yields calibration products (dispersion and $Q$, $\eta_{\rm prop}$, boundary specularity $p$, $E_{\rm thr}$, and $dV/dQ$) that update the nuisance priors in Sec.~\ref{sec:systematics_calibration}. Sensitivity projections are therefore anchored to measured device performance rather than design targets.

\section{Discussion and outlook}
\label{sec:discussion}

\begin{figure}[htp]
    \centering
    \includegraphics[width=0.45\textwidth]{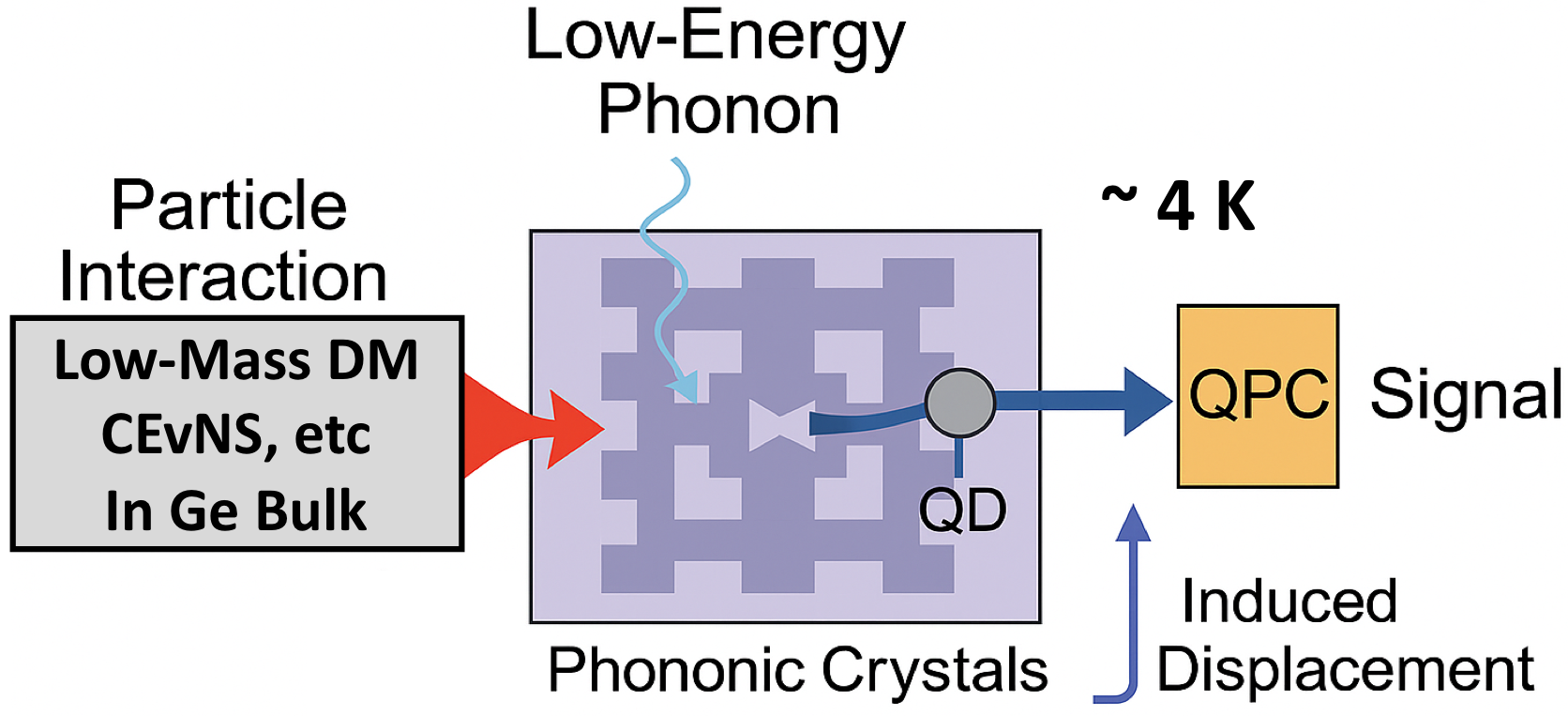}
    \caption{Conceptual block diagram of the GeQuLEP detection chain, from energy deposition and primary-phonon generation to athermal burst guidance in a surface PnC, charge displacement in dipole-defined QDs via the deformation potential, and RF–QPC readout.}
    \label{fig:block_diagram}
\end{figure}

The Ge PnC\,$\to$\,QPC pathway developed here provides a mechanically simple, contact–minimized route to sub-eV sensing that combines \emph{slow-phonon guidance}, \emph{multiplicity-assisted collection}, and \emph{RF-QPC transduction} with timing/PSD discrimination. The operating principle is summarized in Figure~\ref{fig:block_diagram}: rare energy deposits (e.g., DM scattering or CE$\nu$NS) create primary THz phonons that anharmonically down-convert into a burst of ballistic LA/TA modes; surface PnCs steer and slow these modes toward dipole-defined QDs where deformation-potential coupling induces a charge displacement that is read out by a nearby RF-QPC (Secs.~\ref{sec:phonon_physics}--\ref{sec:readout}). Operating at \SI{4}{K} suppresses thermal occupation in the \SI{10}{-}\SI{30}{GHz} analysis band and enables nanosecond timing gates (Sec.~\ref{sec:timing_psd}).

\paragraph{Position in the landscape}
Leading rare-event experiments (e.g., XENONnT~\cite{Aprile2023}, LZ~\cite{LZ2023}, SuperCDMS~\cite{Agnese2018}, EDELWEISS~\cite{Armengaud2019}, CRESST~\cite{CRESSTIII_Si10eV_PRD2023}) have pushed thresholds and backgrounds impressively, yet typical analysis thresholds remain at the eV–keV scale, leaving much of the sub-MeV DM phase space weakly tested~\cite{Aprile2020JCAP,LZ2023,Agnese2014,Armengaud2017}. Semiconductor DM–$e$ detectors (SENSEI~\cite{SENSEI2025PRL}, DAMIC~\cite{DAMICM2023PRL}, skipper-CCD~\cite{AguilarArevalo2019} variants) have probed to few-electron thresholds but still face readout and phononization limitations at ultra-low energies~\cite{Barak2020,AguilarArevalo2019}. By targeting \emph{single-primary-phonon} sensitivity (few\,meV per primary, with multiplicity-assisted collection; Sec.~\ref{sec:collection}), the GeQuLEP architecture opens a complementary channel in which the \emph{first phonon} produced by a sub-eV deposition can be transduced to charge without bulk drift or contacts. The projected sub-meV selection threshold (Sec.~\ref{sec:projected_sensitivity}) moves the turn-on for DM–$e$ scattering into the sub-MeV mass regime for both heavy- and light-mediator benchmarks~\cite{Essig2012,Essig2016,Bloch2017,Knapen2021}. At larger exposures, the solar CE$\nu$NS background becomes the relevant discovery floor~\cite{Billard2014}.

\paragraph{Noise and vibrational environment}
Although dilution refrigerators introduce low-frequency (Hz–kHz) mechanical vibrations, these classical motions have wavelengths and energies far from the GHz, meV-scale acoustic modes that couple efficiently through the deformation potential. The principal risk is \emph{microphonic pickup} in the RF chain, which we mitigate via mechanical isolation, strain-relieved cabling, cold attenuators/isolators, and notch filtering (Sec.~\ref{sec:prototype_backgrounds}). On the sensor side, cryogenic RF-QPCs routinely achieve charge sensitivities $\delta q\!\sim\!10^{-3}e/\sqrt{\mathrm{Hz}}$ and, in optimized reflectometry/multiplexed setups, approaching $10^{-4}e/\sqrt{\mathrm{Hz}}$ within MHz bandwidths~\cite{Schoelkopf1998,Colless2013,GonzalezZalba2015,Cassidy2007,Reilly2007}. These levels are compatible with the induced charges expected from single- or few-phonon bursts in our geometry (Sec.~\ref{sec:readout}).

\paragraph{Prototype realism and backgrounds}
Ultimate sensitivity is set not only by threshold but also by selection efficiency and residual backgrounds. The timing/PSD windowing of Sec.~\ref{sec:timing_psd} is designed to exploit the sub-ns burstiness and spectral concentration of athermal phonons while suppressing the Bose–Einstein tail of thermal phonons at \SI{4}{K}. The calibration plan of Sec.~\ref{sec:systematics_calibration} (heater pulses, narrowband phonon injection, neutron/photo-tags) anchors the acceptance model $\varepsilon(\omega)$ and constrains bulk/surface losses ($\eta_{\rm prop}$, specularity $p$)~\cite{SwartzPohl1989,Klitsner1988}. These \emph{in situ} constraints are essential to convert device-level performance into credible limits (Figure~\ref{fig:proj_sensitivity}) and to quantify systematics in the sub-MeV regime.

\paragraph{Scalability and portability}
The architecture is CMOS-compatible: shallow-impurity QDs, nm-scale PnCs, and planar QPCs are all available within standard nanofab stacks (EBL/DUV, RIE, ALD). Frequency-division or cold combining enables multiplexed RF readout for arrays (Sec.~\ref{sec:roadmap}). Beyond Ge, the approach is portable to other crystalline hosts with (i) low acoustic loss at cryogenic $T$, (ii) strong deformation-potential coupling, and (iii) manufacturable slow-phonon structures—e.g., Si/SiGe, diamond, or sapphire for specialized frequency bands~\cite{Joannopoulos2008,Maldovan2013}. Material choice sets the available band structure, group velocities, and impurity spectra, and can be tailored to specific targets (e.g., dark-photon absorption vs.\ nuclear recoils).

\paragraph{Near-term milestones}
The experimental roadmap (Sec.~\ref{sec:roadmap}) specifies quantitative exit criteria: (i) PnC band center within $\pm\SI{1}{GHz}$ of design and defect-mode $Q\!\gtrsim\!5{\times}10^3$; (ii) baseline charge sensitivity $<\!10^{-3}e/\sqrt{\mathrm{Hz}}$ with stable $dV/dQ$; (iii) selection-corrected thresholds in the $10^{-3}$–$10^{-2}$\,eV band for a \SI{10}{-}\SI{100}{g} module; and (iv) background-free operation to $\sim\!0.1$\,kg$\cdot$yr per module using timing/PSD selections. Meeting these will establish the practical path to kg-scale arrays and the onset of the solar-neutrino background.

\paragraph{Scientific outlook}
If the demonstrated thresholds and selections track design, single-module exposures at the \SI{0.1}{kg\cdot yr} level will probe unexplored DM–$e$ parameter space below the current semiconductor limits~\cite{Essig2012,Barak2020,AguilarArevalo2019}. An array reaching $\mathcal{O}(1)$\,kg$\cdot$yr begins to confront the CE$\nu$NS floor in the \SI{1}{-}\SI{100}{MeV}/$c^2$ regime~\cite{Billard2014}, while simultaneously enabling first detections or stringent bounds on solar CE$\nu$NS with a phonon spectrometer. The same platform naturally supports precision tests of cryogenic phonon transport, impurity physics, and quantum-limited RF sensing—ingredients with impact well beyond the present application.

\section{Conclusion}
\label{sec:conclusion}

We have presented \textbf{GeQuLEP}, a germanium phonon-to-charge platform that functions as a quantum-enhanced \emph{phonon spectrometer} for sub-eV energy deposits. The architecture combines (i) high-purity Ge operated at \SI{4}{K}, (ii) dipole-defined quantum dots that couple strongly to acoustic strain via the deformation potential, (iii) surface phononic-crystal (PnC) structures that \emph{slow}, steer, and confine athermal phonons, and (iv) radio-frequency quantum point-contact (RF-QPC) readout that transduces phonon-driven charge displacement without bulk charge drift or metal contacts across the active volume. This contact-minimized approach turns ballistic athermal phonons into robust information carriers and enables nanosecond timing and GHz-band PSD discrimination.

Our device-level modeling, tied to a calibrated efficiency pipeline, shows that THz primary phonons from rare interactions undergo anharmonic down-conversion into bursts of ballistic modes that are efficiently collected and funneled to the sensing region. Within the QW/PnC, deformation-potential coupling induces charge displacements large enough to generate detectable induced charge on a nearby QPC, with a resonant enhancement in the \SI{10}{-}\SI{30}{GHz} band and measurable response extending to $\gtrsim\!\SI{100}{GHz}$. Under standard halo assumptions, the resulting \emph{selection-corrected} thresholds in the $10^{-3}$--$10^{-2}$\,eV range translate into competitive projected sensitivity to DM--electron scattering from sub-MeV to sub-GeV masses (heavy- and light-mediator benchmarks), with a staged path to kg$\cdot$yr exposures where the solar CE$\nu$NS floor becomes relevant (Sec.~\ref{sec:projected_sensitivity}).

A key outcome of this study is a concrete \emph{calibration and systematics} program (Sec.~\ref{sec:systematics_calibration}) that anchors sensitivity projections to measurable device parameters: baseline noise and effective threshold (heater pulses and narrowband phonon injection), PnC passband and mode $Q$ (dispersion and ringdown), bulk survival and boundary specularity (TOF/echo analyses), and absolute RF-QPC responsivity (charge-step metrology). Together with timing/PSD selections (Sec.~\ref{sec:timing_psd}) and a background program using neutron and photoexcitation tags, these controls bound the leading uncertainties that shape the low-mass turn-on and ensure that limits remain stable against realistic variations in materials and readout.

Looking ahead, the phased roadmap (Sec.~\ref{sec:roadmap}) lays out quantitative exit criteria from cm-scale coupons to \SI{10}{-}\SI{100}{g} modules and, ultimately, kg-scale arrays with multiplexed RF readout. Because the nanofabrication steps (EBL/DUV, RIE, ALD) are CMOS-compatible and the sensing is contact-minimized, the concept is readily portable to other crystalline hosts with low cryogenic loss and strong deformation-potential coupling (e.g., Si/SiGe, diamond), where slow-phonon engineering can be tailored to specific targets. Beyond dark-sector searches and solar CE$\nu$NS, a GeQuLEP-class spectrometer provides a general platform for precision studies of cryogenic phonon transport, impurity-bound quantum states, and quantum-limited RF sensing.

In summary, GeQuLEP merges slow-phonon phononics with charge sensing to realize a practical route to sub-eV detection. If the near-term milestones on threshold, efficiency, and stability are met, the platform will probe unexplored DM--electron parameter space below current semiconductor limits while establishing a scalable, quantum-informed methodology for rare-event detection.

\section*{Credit authorship contribution statement (suggested)}
{\small Conceptualization: D.-M. Mei; Methodology: D.-M. Mei; Investigation: D.-M. Mei, N. Budhathoki, S. A. Panamaldeniya, K.-M. Dong, S. Bhattarai, A. Warren, A. Prem, S. Chhetri; Writing—original draft: D.-M. Mei; Writing—review \& editing: N. Budhathoki, S. A. Panamaldeniya, K.-M. Dong, S. Bhattarai, A. Warren, A. Prem, S. Chhetri ; Supervision: D.-M. Mei}

\section*{Data availability (suggested)}
{\small All data supporting the findings of this study are included in the article; additional information is available from the corresponding author upon reasonable request.}

\section*{Declaration of competing interest (suggested)}
{\small The authors declare that they have no known competing financial interests or personal relationships that could have appeared to influence the work reported in this paper.}

\section*{Acknowledgments}
{\small We acknowledge support from \texttt{NSF OISE 1743790, NSF PHYS 2310027, NSF OIA 2437416, DOE DE-SC0024519, DE-SC0004768 } and a research center supported by the State of South Dakota.}

\bibliographystyle{elsarticle-num}
\bibliography{refs}

\end{document}